\documentstyle[12pt,psfig,subfigure,amstex]{l-aa} % LaTeX A&A  Standard Fonts
\begin{document}
   \thesaurus{00 (12.03.3;12.12.1)}
   \title{The ESO-Sculptor Faint Galaxy Redshift Survey:
	   The Photometric Sample 
   \thanks{Based on observations collected at the European Southern Observatory, 
    La Silla, Chile}}

  \author{S. Arnouts \inst{1}
       \and 
         V. de Lapparent \inst{1}
       \and
         G. Mathez   \inst{3}
       \and
         A. Mazure   \inst{4}
       \and
         Y. Mellier  \inst{1,3}
       \and
         E. Bertin \inst{1,2}
       \and
         A. Kruszewski \inst{5}}
    \offprints{S. Arnouts}
     
   \institute{CNRS, Institut d'Astrophysique de Paris,
	       98 bis, Boulevard Arago,
	       F-75014 Paris,
	       France
         \and  
           European Southern Observatory, Casilla 19001, Santiago, Chile
         \and 
           Observatoire Midi-Pyr\'en\'ees, Laboratoire d'Astrophysique de
         Toulouse, URA 285, 14, avenue Edouard Belin, F-31400 Toulouse, France
         \and 
           Laboratoire d'Astronomie Spatiale, B.P. No. 8, F-13376 Marseille Cedex 
         12, France
         \and
           Warsaw University Observatory, Al. Ujazdowskie 4, PL-00-478 Warsaw,
           Poland}
             
   \date{Received August 5; accepted October 16,1995}

   \maketitle 
%
% --------------------------------------------------------------
%
\begin{abstract}

We present the photometric sample of a faint galaxy survey carried out
in the southern hemisphere, using CCDs on the 3.60m and NTT-3.5m
telescopes at La Silla (ESO). The survey area is a continuous strip of
$0.2^{\circ} \times 1.53^{\circ}$ located at high galactic latitude
($b^{II} \sim -83^{\circ}$) in the Sculptor constellation.  The
photometric survey provides total magnitudes in the bands B, V
(Johnson) and R (Cousins) to limiting magnitudes of 24.5, 24.0, 23.5
respectively. To these limits, the catalog contains about 9500, 12150,
13000 galaxies in B, V, R bands respectively and is the first large
digital multi-colour photometric catalog at this depth.  This
photometric survey also provides the entry catalog for a fully-sampled
redshift survey of $\sim$ 700 galaxies with $R\le 20.5$ (Bellanger et
al. 1995a).\\ 
In this paper, we describe the photometric observations
and the steps used in the data reduction.  The analysis of objects and
the star-galaxy separation with a neural network are performed using
SExtractor, a new photometric software developed by E. Bertin (1996).
By application of SExtractor to simulated frames and comparison of
multiple measurements, we estimate that the photometric accuracy of
our catalog is $\sim$ 0.05$^m$ for $R \le 22$.  Then, we use a method
to obtain a homogeneous photometric scale over the whole survey using
the overlapping regions of neighbouring CCDs.  The differential galaxy
number counts in B, V, R are in good agreement with previously
published CCD studies and confirm the evidence for significant
evolution at faint magnitudes as compared to a standard non evolving
model (by factors 3.6, 2.6, 2.1).  The galaxy colour distributions
B$-$R, B$-$V of our sample show a blueing trend of $\sim 0.5^m$
between 21 $<$ R $<$ 23.5 in contrast to the V$-$R colour distribution
where no significant evolution is observed.  
\end{abstract}

\section{INTRODUCTION}

During the past decade, the photometric and spectroscopic surveys have
allowed to improve our knowledge of galaxy formation and evolution.
Since the 1980's, the new technology based on CCD detectors has
improved the photometric efficiency (sensitivity, linearity, high
dynamic range,...) as compared to photographic plates. Although the
photographic plates are un-rivaled in their ability to cover large
areas of the sky, the large improvements brought by digital surveys,
both in the photometric accuracy and in the faint limiting magnitude
which can be reached, have allowed important new insight into the
properties and the evolution of the different galaxy populations.
Several deep digital surveys (Tyson 1988, Lilly et al. 1991, Metcalfe
1991, 1995, Driver et al. 1994, Smail et al.  1995) were performed in
different regions of the sky (with a typical field size $\le$ 0.012
deg$^2$) at faint (B$\le$ 25) or very faint magnitudes (B$\le$ 27.5).
First, these surveys show that the number-counts of galaxies at B
$\simeq$ 22 are in excess with respect to a non-evolving model and the
disagreement increases with the apparent magnitude. Second, the galaxy
colour distributions become significantly bluer at fainter
magnitudes. \\ 
Several models have been elaborated based on the
cosmological parameters (H$_0$, q$_0$) and the luminosity function
parameters ($L^{\star}, \Phi^{\star}, \alpha$) allowing a good fit to
the photometric data (see Koo Kron 1992 for a review).  Models with
pure luminosity evolution where $L^{\star}$ evolves with look-back
time (Bruzual 1983, Guiderdoni \& Rocca-Volmerange 1990, Yoshii \&
Takahara 1988) predict that the tail of the redshift distribution of
very faint galaxies should be extended towards high redshifts.  This
is not observed in the recent redshift surveys to $B\le$ 24 (Colless
et al. 1990, 1993, Cowie et al. 1991, Lilly et al. 1991, Tresse et
al. 1994 and Glazebrook et al. 1995a), which are in good agreement
with the redshift distribution expected for a non-evolving model.
Models with number-density evolution (Rocca-Volmerange \& Guiderdoni
1990, Broadhurst et al. 1992) are based on a population of dwarf
galaxies at z$\sim$ 0.3 which would have merged into brighter galaxies
by z$\simeq$0.  However, these models are difficult to reconcile with
both the recent observations of weak clustering in the correlation
function of faint galaxies (Efstathiou et al. 1991 and Roche et
al. 1993) and with the physical mechanisms for merging (Ostriker 1990,
Dalcanton 1993).  The most recent models use a new estimation of the
slope $\alpha$ of the local luminosity function (Koo et al. 1993,
Driver et al. 1994) which is assumed to increase from $\alpha \simeq
-1.1$ to $\alpha \simeq -1.8$ by the presence of a large number of
dwarf galaxies (dI,dE). This model is supported by the recent
observations of the Medium Deep Survey with the HST to I=22
(Glazebrook et al. 1994b), where the counts of morphologically normal
galaxies are well fitted by a non-evolving model and where a large
excess of Irregular and Peculiar galaxies is detected which could
contribute to the excess of blue galaxies.

Faint photometric and spectroscopic surveys also provide maps of the
distribution of galaxies in three dimensions. The nearby surveys show
that galaxies are distributed within sharp walls delineating voids
with diameters between 20 and 50$h^{-1}$ Mpc (de Lapparent et
al. 1986, Geller \& Huchra 1989, da Costa et al. 1994) (where $H_0$ =
100 $h\ km.s^{-1}.Mpc^{-1}$).  Very deep pencil beam surveys were
obtained in particular directions of the sky (Broadhurst et al. 1988,
Colless et al. 1990, 1993, Cowie et al. 1991, Lilly et al. 1991,
Tresse et al. 1994 and Glazebrook et al. 1995a), and in some of these
an apparent periodicity on scales of $\sim$ 128$h^{-1}$ Mpc has been
detected (Broadhurst et al. 1990).  At these depths (z $\simeq$
0.3$-$0.5), the large amount of time required to obtain the redshift
distribution for a complete magnitude-limited sample constrains
observations to narrow solid angles. Although the existing deep
pencil-beam probes are adequate for establishing the evolutionary
history of galaxies, biases caused by sparse sampling may affect the
data when used to study the large scale structures (de Lapparent et
al. 1991, Ramella et al.  1992).

In this context, a deep redshift survey near the southern galactic
pole was started with the main goal to characterize the large-scale
structure at large distances (de Lapparent et al. 1993).  The
spectroscopic survey covers a continuous solid angle of 0.3 deg$^2$
and contains $\sim$ 700 galaxies with R $\le$ 20.5 (i.e. B $\le$ 22).
The median redshift is at z $\simeq$ 0.3.  The entry photometric
catalogue for the redshift survey was obtained by observing in the B,
V, R photometric bands up to 24.5, 24.0, 23.5 respectively in a longer
strip of 0.4 deg$^2$ (this area was not fully covered by spectroscopic
observations).  The photometric data provide an adequate sample for
measuring with high confidence level the galaxy number counts and the
distribution of galaxy colours.  The description of the spectroscopic
sample of the ESO-Sculptor survey is given in Bellanger et al. (1995a)
and the first results about the properties of the large-scale
structure are shown in Bellanger \& de Lapparent (1995b).

Here we describe in detail the procedures used in the reduction and
analysis of the photometric sample.  The paper is organized as
follows. In section 2, we describe the photometric
observations. Sections 3 and 4 outline the data reduction procedures
and analyses. In sections 5, 6 and 7 we discuss the transformation
into astronomical coordinates, the photometric calibration of fields
and the magnitude transformations into the Johnson-Cousins standard
system.  In section 8 we present the method used to match the
photometry over the whole survey in each band. In section 9 we show
the first results on the star colour distributions, the galaxy colours
and the galaxy number-counts.  Finally, in section 10 we summarize the
major steps of our photometry and we present the scientific
perspectives for the near future.

\section{OBSERVATIONS}

This survey was performed in the context of an ESO-key programme, thus
guaranteeing the necessary observing time to perform the full
programme. The observations began in 1989 at the 3.60m telescope and
were transferred to the NTT-3.5m in 1991.  At a pace of two observing
runs per year, the full time allocation was consumed by the fall 1995.
The observing strategy is designed to make optimal use of telescope
time.  The great advantage of the multi-mode instruments used (EMMI at
the NTT, D'Odorico 1990, Dekker et al. 1991, and EFOSC at the 3.6m,
Buzzoni et al. 1984) is the possibility to switch between
spectroscopic and photometric observations during the course of the
same night, using the same instrument.  It allows one to adapt to
variable weather conditions : when the conditions are photometric with
good seeing quality, priority is given to imaging; if the weather
degrades, one can switch to the spectroscopic mode by simple rotation
of two wheels placing an aperture mask and a grism into the optical
path.  Because the number of observing nights is spread over many
runs, the imaging data of a given run can be reduced in preparation
for the subsequent run, thus yielding finding charts of the galaxies
with $ R \le 20.5 $ to be observed spectroscopically.  \\ 
During each
run, we observe our main survey region in first priority, and two
auxiliary regions at the beginning and end of the night when the
airmass for the main survey is large.  Here we only report on the data
for the main survey region. The data in the auxiliary fields will be
used as control samples.  The position, galactic latitude and solid
angle of the main field are given in Table 1.  In this region of the
sky, we acquired 25-30 frames in each of the B, V, R bands with the
3.6m telescope.  These data cover the right ascension range $0^h 19^m
00^s < R.A._{2000} < 0^h 21^m 24^s$, and correspond to an area of
$\sim 0.53^{\degr} (R.A.) \times 0.24^{\degr} (Dec)$.  With the NTT
telescope, we acquired 20-25 frames with $0^h 21^m 24^s < R.A._{2000}
< 0^h 26^m 00^s$ ($\sim 1.00^{\degr} (R.A.) \times 0.24^{\degr}
(Dec)$).  The $0.24^{\degr}$ width in declination is covered by 3
EFOSC fields and 2 EMMI fields.
%--------------------------------------------------------
\begin{table}
  \label{tabprob}
  \caption{Survey characteristics}
  \begin{tabular}{ccc}
  \hline
  center      & $b^{II}$ & Total area \\
 R.A., DEC.&  (deg)   & (deg$^2$) \\
  \hline
$0^h 22^m 30^s$,$-30^{\degr} 06^{\arcmin}$&-83&0.24$\times$1.53 \\ 
  \hline
  \end{tabular}
\end{table}
%-------------------------------------------------------- 
% 
\subsection {Instrumentation} 

Because we performed the photometric observations over a period of six
years, different CCD cameras were used over the years to follow the
improvement in CCD technology, so that the size and the
characteristics of the individual CCD frames vary across the mosaic
(made of typically 50 frames in each band).  We summarize the
principal CCD characteristics in Table 2.  The first column gives the
telescope and instrument used with the different channel for EMMI on
the NTT (blue imaging channel: BIMG, and red imaging channel: RIMG).
The useful area is the used part of the CCD which differs for large
CCDs (LORAL, TEKTRONIX) from the total area because of vignetting.
%--------------------------------------------------------
\begin{table*}
  \label{tabccd}
  \caption{CCD characteristics and Photometric observations}
  \begin{tabular}{cccccccc}
  \hline
 Telescope/instrument & CCD & Pix size & Useful area & Period & \multicolumn{3} {c} {Average zero-points} \\
                      &     & (''/pix) & (arcmin$^2$)&        &  B & V & R               \\
  \hline
  3.60m/EFOSC &RCA\#8 &0.675&3.7$\times$5.7& 43-44 & 23.70$\pm$0.03 & 24.24$\pm$0.04 & 24.21$\pm$0.02  \\
  3.60m/EFOSC &RCA\#8 & -   & -            & 45-46 & 23.41$\pm$0.01 & 24.06$\pm$0.03 & 24.03$\pm$0.02   \\
  NTT/EMMI (RIMG)&THX\#18&0.44 &7.5$\times$7.5& 49-50 &  & 24.30$\pm$0.02 & 24.65$\pm$0.03 \\
  NTT/EMMI (RIMG)&LOR\#34&0.35 &9.3$\times$8.8& 52 & 24.69$\pm$0.02$^1$ &  & 25.14$\pm$0.02  \\
  NTT/EMMI (RIMG)&TEK\#36&0.27 &9.1$\times$9.1& 54 &  & 25.40$\pm$0.02 &             \\
  NTT/EMMI (BIMG)&TEK\#31&0.37 &6.2$\times$6.2& 52 &  24.69$\pm$0.02 & &    \\
  NTT/EMMI (BIMG)&TEK\#31& - & -              & 54 &  24.26$\pm$0.01 & &    \\
  \hline
  \end{tabular}
\end{table*}
%--------------------------------------------------------

\subsection{Observations}

At the 3.60m telescope with EFOSC, the photometric observations were
performed in a range of seeing between 1.10 and 1.65 arcsec. When the
FWHM is $<$ 2 pixels, the profile of unresolved objects is poorly
sampled, which degrades the star/galaxy separation. With EFOSC, we
observed the photometric fields with airmasses between 1.0 and 1.4,
and the exposure times for the B, V, R bands are respectively 30 , 25,
20 minutes.  The majority of the fields were obtained in only 1
exposure in each band.  Bright stars (R $\le$ 18 mag) are thus
saturated.  Since 1991, we have observed at the NTT with an average
seeing of $\sim$ 1.0 arcsec.  As the pixel size of the CCDs is
smaller, we have a better sampling of the p.s.f. and therefore our
star/galaxy separation is limited by seeing.  The airmasses are also
in the range 1.0 - 1.4. The exposure times for the B, V, R bands were
respectively 25, 20, 15 minutes.  To avoid saturation of the CCDs by
bright stars, the images are summed from 2 or 3 exposures.  The
photometric strategy adopted to probe the different strips is to
obtain a mosaic of CCD frames regularly offset by $9 \over 10$ of the
CCD size providing many overlaps for subsequent checks of the
photometry.  In Table 2, we summarize the characteristics of the
photometric observations during the different runs. For each run, the
photometric zero-points (apparent magnitude of a star with an absolute
magnitude $M = 0.$ and null color term exposed during 1 second, as
defined in Equation (1)) are given.
\section {DATA PRE-REDUCTION}

Before the photometric analysis, the raw data are reduced using the
MIDAS environment in the following steps:\\ 
bias subtraction, skimming
subtraction (for RCA CCD \#8), flat-fielding and cosmic-ray removal.
\subsection{Bias subtraction}

A ``master'' bias frame is obtained by averaging over $\sim$ 50 frames
from which cosmic ray events have been removed.  As the bias level
varies with time, the bias subtraction for each scientific frame is
performed in two steps : \\ 
(1) Subtraction of the master bias frame\\
(2) Subtraction of the mean difference between the overscan of the
scientific frame and that of the master bias frame.
\subsection{Skimming subtraction}

With RCA CCD \#8, the raw CCD frames have a skimming pattern,
characterized by systematic column intensity offsets across the entire
CCD.  This pattern varies in intensity at low illumination levels and
becomes constant above $\sim$ 1000 adu, well below the sky level of
all our scientific frames but above that for the photometric
calibration frames.  Because the skimming offsets are additive biases,
we can calculate them by using flat-field frames with different
illumination levels. Scaling and subtraction of two flat-field frames
with different exposure times yield a preliminary skimming frame.  As
the skimming feature is stable along each column, the signal to noise
is improved by replacing each pixel of a column by the mean value
along the column.  The mean level of the entire skimming frame is then
adjusted to be zero.  A set of skimming frames derived from pairs of
flat-field frames with increasing illumination levels are
calculated. Then for each science frame, the appropriate skimming
frame can be subtracted.  The skimming subtraction is important for
the calibration frames where the sky level is very low. To preserve
the quality of our photometry, the calibration frames for which we
cannot adequately subtract the skimming pattern are rejected.
\subsection{Flat-fielding}

The flat-field frame provides a map of the sensitivity variations over
the CCD chip.  This map depends on the spectrum of the incoming
light. As the flat-field pattern is stable from night to night, we
obtain for each observing run a ``super flat-field '' in each filter
band by doing a median filtering of all the scientific frames obtained
during the run.  The median filtering removes the objects from the
images and yields the large scale sensitivity variations. This is the
best flat-field frame which can be obtained, because it is derived
from the sky on the science frames themselves.  The number of frames
must be sufficient to create a high signal to noise final flat-field
frame. In practice, because our fields are sparsely populated, the
super flat-field results from the median filtering of 5 to 20 science
frames (depending on the observing run).  This flat-field is
normalized and is divided into each data frame.  The large-scale
residual variations in the background of each flat-fielded frame are
$<$1 \%.
\subsection{Summation of multiple exposures}

After flat-fielding, we align multiple exposures of identical fields
using several unresolved objects, sum the individual exposures and
calculate the mean airmass for the final frame.
\subsection{Cosmic ray removal}

To remove cosmic rays from the summed frame, we apply a filtering
algorithm kindly provided by P. Leisy.  Each pixel is compared with
the mean of the $5 \times 5$ neighbouring pixels. If the pixel value
differs by more than $5\sigma$ from the mean, it is replaced by the
mean value. The value of $5\sigma$ is large enough to prevent the
subtraction of real objects.
\section{PHOTOMETRIC ANALYSIS}
 This section briefly describes the photometric software used in the
data analysis (SExtractor : Source Extraction software).  A complete
description is given in Bertin \& Arnouts (1996, hereafter BA96).
\subsection{Description of photometric package}
 The photometric analysis is done in four steps: 
\subsubsection{determination of the sky background}
The background map is derived by binning the frame into large meshes
(32 $\times$ 32 or 64 $\times$ 64 pixels), and removing the possible
overestimations due to bright objects using a $\kappa$-sigma clipping
algorithm.  This sky background map is subtracted from each frame.

\subsubsection{Detection of objects}
The detection algorithm determines a group of connected pixels above a
given threshold. A minimum number of connected pixels is chosen in
order to avoid spurious objects (bad pixels, un-removed cosmic
rays,...). Convolution with a gaussian with a FWHM close to the seeing
improves the detection of objects by decreasing the background noise.
The typical threshold used is $1.2\sigma$ above the background value (
$\sigma$ is measured from the unsmoothed background ) corresponding to
a surface brightness equal to 27 mag.arcsec$^{-2}$ in B, 26.5
mag.arcsec$^{-2}$ in V and 26 mag.arcsec$^{-2}$ in R.  When such a low
threshold is used, large spurious faint objects can appear in the
wings of objects with shallow profiles. This effect is seen around
elliptical galaxies or bright stars where the local background noise
increases and can exceed the detection threshold.  A cleaning
procedure is therefore applied to check if the detected objects are
real.  For the faint objects in the vicinity of bright objects, a new
estimation of the local background is obtained by assuming that the
dominant central object has large gaussian wings. If the mean surface
brightness of the faint object is lower than the calculated local
threshold, the object is rejected.
\subsubsection{Deblending of neighbouring objects:}
Each detection of a connected set of pixels is processed through a deblending 
algorithm based on  multi-thresholding as described in  BA96. 
\subsubsection{Photometry}
Three kinds of magnitudes are used: isophotal, ``corrected isophotal''and 
``adaptive aperture''.

The ``corrected isophotal'' retrieves the lost flux in the isophotal
magnitude by assuming that the wings of objects outside the limiting
isophote are nearly gaussian (see Maddox et al. 1990a). This
estimation of the ``total'' magnitude could be improved by assuming
that the profile of objects follows the more realistic Moffat
profile. However, if the value of the threshold is low, the gaussian
profile provides an adequate correction as demonstrated by the tests
on simulated frames (see section 4.2).

The ``adaptive aperture'' magnitude is the best estimation of the
``total'' magnitude.  The algorithm is similar to the ``first-moment''
measure designed by Kron (1980).  This magnitude is calculated in two
steps:

 (1) The object's light distribution above the isophotal threshold is
used to measure an isophotal elliptical aperture characterized by the
elongation $\epsilon$ and position angle $\theta$.
 
 (2) The first moment $r_1$ is calculated in an aperture 2 $\times$
larger than the isophotal aperture in order to reach the light profile
information below the isophotal threshold. \\ 
The first moment is used
to define the ``adaptive aperture'' of radius $k r_1$ inside which we
measure the ``total'' magnitude.  The principal axes of each object
are defined by $\epsilon k r_1 $ and $ k r_1/\epsilon$.  Kron (1980)
uses a circular aperture of radius $k r_1$ with $k=$2 which measures
90\% of the total flux from the objects. To converge near the
``total'' magnitude we can increase the $k$ value, but a compromise
must be found between the added measured flux and the increasing noise
in larger apertures.  As Metcalfe et al. (1991), we choose $k=$2.5,
yielding 94\% of the total flux inside the adaptive aperture (this
value was calculated using simulated frames with a large variety of
galaxy profiles).  In contrast to the isophotal magnitude which
operates at fixed signal-to-noise, the ``adaptive aperture'' magnitude
may be determined at very low signal-to-noise. Sometimes for faint
objects, $r_1$ may converge to erroneously small apertures.  We
therefore constrain the apertures to a minimum value of $R_{min} \sim$
3.5 $\sigma_{ab}$ (where $\sigma_{ab}$ is the mean standard deviation
of the bivariate gaussian profile defined by the second order moments
of the object profile (BA96)).

\subsection{Simulations}
To test the quality of the photometry for objects with different
magnitudes, we use simulated frames provided by E. Bertin (BA96) and
generated according to the following scheme.  A sky background is
defined by a sky surface brightness ($\mu_R = 21$ mag/$arcsec^2$), and
Poisson noise is added according to a gaussian distribution as
observed on real CCD frames.  A set of stars with a Moffat profile
(Moffat 1969) are generated.  Galaxies are produced with a large
variety of shapes and sizes.  The pixel size and seeing disk are
adjusted to resemble as much as possible the real frames.  The frames
are defined through the R band corresponding to the band used here for
the selection of the spectroscopic sample and for the star/galaxy
separation.  The results of the tests on several simulated images are
based on $\sim$ 6000 galaxies and $\sim$ 600 stars in total.
\subsubsection{Test of photometric accuracy}

\begin{figure}
\centerline{\psfig{figure=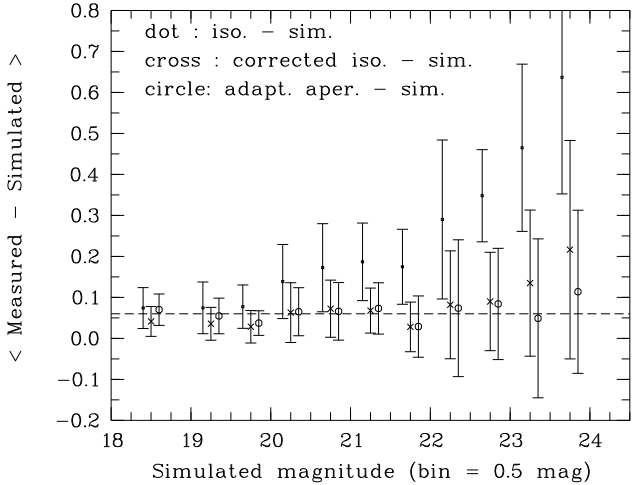,angle=-90,height=7.cm}}
       \caption[]{ Mean difference in 0.5$^m$ bins between the measured magnitudes 
and the true magnitudes for galaxies in simulated R frames. The symbols for
the different
 magnitudes used are given inside the graph. The error bars show the r.m.s 
 scatter around the mean. The dashed line represents the expected
 position of magnitude difference for the  ``adaptive aperture'' magnitude 
 corresponding to the 94 \% enclosed flux.}
            \label{difmagg}
    \end{figure}

In Fig. ~\ref{difmagg}, we compare the mean difference between the
different measured magnitudes and the true magnitude for the
galaxies. The mean difference is calculated in bins of 0.5$^m$ of the
true magnitude.  The error bars represent the r.m.s scatter around the
mean.\\ 
As expected, the isophotal magnitude which measures the flux
inside the defined isophote looses the flux in the wings outside this
isophote.  Fig. ~\ref{difmagg} shows that an increasing fraction of
flux is lost for fainter magnitudes.  The ``corrected isophotal''
magnitude (defined in section 4.1.4) provides a significant
improvement. \\ 
Fig. ~\ref{difmagg} also confirms that the ``adaptive
aperture'' magnitude (Kron magnitude) measures $\sim$ 94\% of the flux
of objects over the entire magnitude range (the systematic offset of
0.$^m$06 is indicated by the dashed line in Fig.~\ref{difmagg}).  To
reach the ``total'' magnitude, we then subtract a constant value of
0.$^m$06 to all magnitudes.  The error bars in Fig.~\ref{difmagg} are
$\sim$ 0.05$^m$ for galaxies with R $\le$ 21., and up to $\sim$
0.2$^m$ at fainter magnitudes.  The interest of the ``adaptive
aperture'' magnitude over the isophotal magnitude originates in it
insensitivity to seeing and redshift (Kron 1980).\\

\begin{figure}
\centerline{\psfig{figure=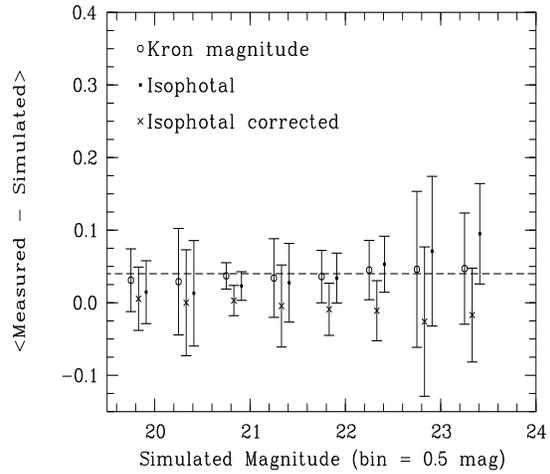,angle=-90,width=8.5cm,height=7.cm}}
       \caption[]{ same as Fig.~\ref{difmagg} for stars. The dashed line represents 
           the flux measured inside a radius defined as 3.5$\times \sigma_{a}$,
  where $\sigma_{a}$ is derived  from stars with high signal-to-noise  
 (as described in text).}
            \label{difmags}
    \end{figure}

Because a large fraction of the detected objects in our survey are
stars, we also compare in Fig.~\ref{difmags} the different magnitudes
for the simulated stars.  Even when a gaussian profile is used to
correct the flux lost in the wings of stellar objects with Moffat
profile, the ``corrected isophotal'' is very close to the total
magnitude for R$\le$ 23.5. \\ 
Because the fraction of flux measured in
a fixed aperture is constant for stars, we check the reliability of
the second order moments by comparing the ``adaptive'' magnitudes to
the ``aperture'' magnitudes.  The aperture radius is defined by
3.5$\times \sigma_{a}$ where $\sigma_{a}$ is the second order moment
along the major axis, estimated by using unsatured stars with high
signal-to-noise (typically 19$^m \le R \le 22^m$). The lost flux is
stable in the full magnitude range and is close to 0.04$^m$.  This
value is in good agreement with the ``adaptive'' aperture obtained by
using a minimum radius defined as 3.5 $\sigma_{ab}$ (where
$\sigma_{ab} = \sqrt{\sigma_a.\sigma_b}$ is calculated for each
object). We can then conclude that the second order moments can be
reliably estimated down to very low signal-to-noise ratios and the
``adaptive'' magnitude is a good measure of the total magnitude.
      
Despite the reliability of ``adaptive'' magnitude for stars, we use
the ``corrected isophotal'' magnitude as being the best magnitude to
estimate the ``total'' magnitude .  Because, we make the star/galaxy
separation only for R $\le$ 22$^m$, we use the ``corrected isophotal''
magnitude up to this cut-off for the stellar sample and the
``adaptive'' magnitude for the galaxy sample. At fainter magnitudes,
no classification is done, thus we use the ``adaptive'' magnitude for
all objects.
%--------------------------------------------------------
%--------------------------------------------------------
%--------------------------------------------------------
%--------------------------------------------------------
\subsubsection{Conclusion}
The results of the tests applied to the simulated images show that : 

 (1) For galaxies, the ``adaptive aperture'' magnitude is a robust
estimate of the ``total magnitude'' (after correction for an offset of
0.06$^m$) down to R $\le$ 24.0. The uncertainty in the measure is
close to 0.05$^m$ up to R$\le$21 and increases to 0.2$^m$ up to
R$\le$23.5.  However, this is an aperture magnitude and it is
sensitive to crowding by close neighbours. Therefore, when an object
has a neighbours closer than 2 isophotal radii, we use by default the
``corrected isophotal'' magnitude.

 (2) For stars, the ``corrected isophotal'' magnitude is a reliable
estimate of the ``total magnitude'' for R $\le$ 23.5.

 (3) Because we measure ``total'' magnitudes, the colours can be
calculated as the difference between the ``total'' magnitudes in each
passband. The colour error can be estimated as the quadratic sum of
the error in each band. For objects brighter than R $\le$ 22, the
colour error is $\simeq$ 0.07$^m$ and at fainter magnitudes the colour
error is $\simeq$ 0.28$^m$.

\subsection{Star-galaxy separation}
\begin{figure}
      
\centerline{\psfig{figure=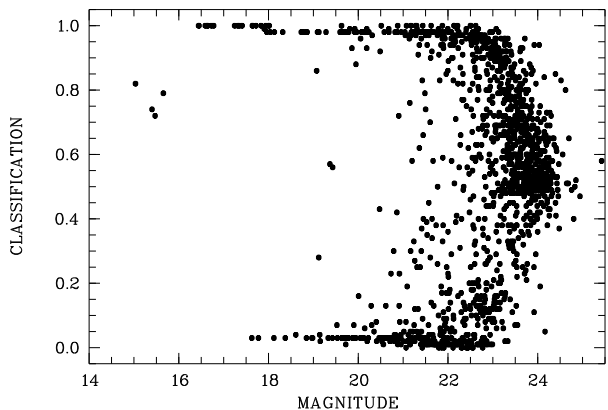,angle=-90,width=9.cm}}
       \caption[]{ Star/galaxy separation for a science frame observed in the
  R band with a seeing of 0.9 arcsec. The ordinate  shows the output parameter of neural 
network , the ``stellarity index '' which provides an estimate of the confidence 
 level in the classification  of each object as a star (CLASS = 1) or a galaxy 
 (CLASS = 0).}
            \label{stargal}
    \end{figure}

To build the spectroscopic catalogue, it is important to exclude the
stellar objects. The star/galaxy separation is performed using the
neural network included in the photometric software SExtractor. For a
complete description of the principle, the training (with using
simulated frames containing stars and galaxies), and various tests of
this neural network, the reader can refer to BA96.  Here, we describe
the input and output parameters used for the separation.  For each
image, the classifier works in a ten-parameter space containing 8
isophotal areas defined by dividing the peak intensity in 8 levels
equally spaced in logarithmic scale, the central peak intensity, and
one control-parameter which defines the fuzziness of the frame and is
chosen to be the seeing (in pixels).  This set of input parameters is
shown to provide an ``optimal'' description of the characteristics of
each image (BA96).  The neural network provides an output parameter
defined as a ``stellarity-index''. Because the p.s.f. of a frame is
the parameter which determines the quality of the performance of the
neural network, SExtractor estimates on each frame the FWHM of the
p.s.f. using the unsatured bright stars and uses it for the neural
network.  The ``stellarity-index'' output parameter is a measure of
the confidence level in the classification of each image.  The value
varies between 1 for stars and 0 for galaxies.  In Fig.~\ref{stargal},
we show the stellarity index as a function of magnitude for a R
frame. For our 50 R frames, a reliable classification can be done for
R $\le$ 22 with a success rate for the galaxies close to 95\%
(BA96). A fainter magnitude limit can be reached for several frames
with good seeing quality (such a frame is shown in
Fig. ~\ref{stargal}).  To be sure that galaxies are not mis-classified
as stars, we define as stars all objects with a stellarity index
greater than 0.8.  The corresponding magnitude limit for
classification is 1.5 magnitude deeper than our spectroscopic limit of
R=20.5, and thus guarantees that our spectroscopic sub-sample is
poorly biased by mis-classified objects. \\ 
Fig.~\ref{stargal} shows
that at faint magnitudes R $\ge$ 23, the stellarity index approaches
0.5 because no distinction can be made between the two classes when
the profiles are dominated by the seeing disk. On the other hand, at
bright magnitudes, for the largely saturated stars (R $\le$ 16) the
stellarity index may drop below 0.8.  \\ 
When we started the
spectroscopic observations, the photometric software by E. Bertin was
not avalaible. We thus used INVENTORY, the photometric software
developed by A. Kruszweski (West \& Kruszewski (1981)) and avalaible
in the MIDAS environment.  INVENTORY provides a reliable star-galaxy
separation for R$\le$21 and was used to generate the entry catalogue
for the first half of the spectroscopic observations. The good success
rate in the separation is confirmed by only 15 spectra of stars
observed out of a total number of 521 reduced spectra (i.e. $\le$ 3\%)
(Bellanger et al. 1995a).
        
%--------------------------------------------------------
%--------------------------------------------------------

\section{ASTRONOMICAL COORDINATES}

To obtain precise astrometry for the detected objects in our CCD
frames, the regions of the SERC-J photographic plates containing our
survey were digitized at the MAMA (``Machine Automatique \`a Mesurer
pour l'Astronomie '' at Paris Observatory) which is developed and
operated by CNRS/INSU (Institut National des Sciences de l'Univers).

The astrometry is performed in two steps.\\ 

(1) By alignment of the images of the digitized plates onto our CCD
frames (using existing MIDAS commands applied to objects in the
field), we are able to derive the coordinate transformation equations
from CCD to photographic plate. There is one set of transformation
equation for each CCD frame. \\ 

(2)The transformation equations from photographic plate to equatorial
coordinates are provided by the MAMA facility as FORTRAN programs. The
astrometric reduction uses the PPM catalogue (Roeser \& Bastian
1991). \\ 

(3) Combination of the two transformations yields for each CCD frame
the transformation equations from pixels to equatorial coordinates.
We use a set of MIDAS procedures which were written with the goal to
perform the transformation routinely for the numerous fields of the
programme (Revenu and de Lapparent 1992).\\ 

\begin{figure}     
\centerline{\psfig{figure=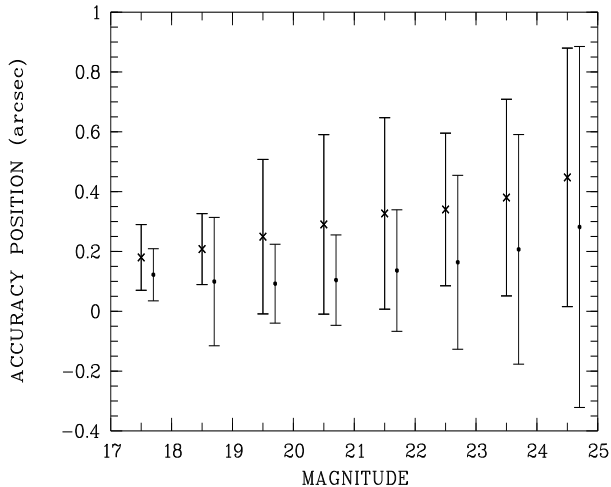,angle=-90,height=7.5cm,width=9.5cm}}
      \caption[]{ Mean difference in arcsec in the astrometric position of 
 common objects in CCD overlaps as a function of R magnitude in bin sizes of 1 mag.
 The data for the NTT and 3m60 telescopes are represented by dots and crosses respectively.
 The error bars show the 1 $\sigma$ r.m.s. dispersion.}
            \label{medastrom}
    \end{figure}

The overlapping regions of neighbouring CCD frames allow us to
estimate the internal astrometric consistency.  In
Fig.~\ref{medastrom}, we plot the median position difference defined
as $\sqrt{cos^{2}(DEC_0) \times (RA_0 - RA_1)^{2} + (DEC_0 -
DEC_1)^{2}}$ in bin sizes of 1 mag.  The NTT observations are
represented by dots and those from the 3m60 by crosses.  The error
bars are the 1 $\sigma$ dispersion measured in each magnitude bin.
For the NTT, the differences of median position vary from 0.1 $\arcsec
\pm 0.1 \arcsec$ for R$\le$ 22 to 0.2 $\arcsec \pm 0.3 \arcsec$ for
R$\le$24, and from 0.2 $\arcsec \pm 0.2 \arcsec$ for R$\le$ 20 to 0.4
$\arcsec \pm 0.3 \arcsec$ for R$\le$24 with the 3m60.  The better
accuracy for the NTT fields is due to the better sampling (smaller
pixel size) which improves the measurement of the central position of
the objects: the seeing disk is spread over $\sim$ 3 pixels for NTT
frames, whereas it is spread over only $\sim$ 2 pixels for 3m60
frames.
%--------------------------------------------------------
%--------------------------------------------------------
\section{CALIBRATION}
 To calibrate our CCD magnitudes into the Johnson B, V and Cousins R standard system
 (hereafter B, V, R),
we observed the faint standard star sequence NGC 300 from Graham (1981) and 
 the bright standard star sequences MARKA, NTPHE, PG0231, SA92 from Landolt (1992). 
 Several sequences were  observed each night in order to obtain a zero-point for 
 each night. 
 We use the following  linear transformation equations to convert instrumental
magnitudes $m_{ccd}$ into standard magnitudes $M_{std}$ :
%--------------------------------------------------------
\begin{equation}
 M_{std} = m_{ccd} - A_M \times sec\xi + k_M \times COLOUR  + C_M 
\end{equation}
%--------------------------------------------------------

where $M_{std}$ define the standard magnitudes and m$_{ccd}$ define
the instrumental magnitudes (in adu.sec$^{-1}$) through respectively
the three filters B,V,R.  The colour term COLOUR is the standard
colour provided by Graham or Landolt B$-$V or V$-$R.  The instrumental
CCD magnitude of standard stars is calculated as the ``corrected
isophotal'' magnitude (see section 4.2.1), thus yielding a good
estimation of the total magnitude.\\ 
The extinction coefficients $A_M$
for the different filters B, V, R derive from the La Silla extinction
curve provided in the ESO manual (Schwarz \& Melnick 1993).\\ 

\begin{figure}
      
\centerline{\psfig{figure=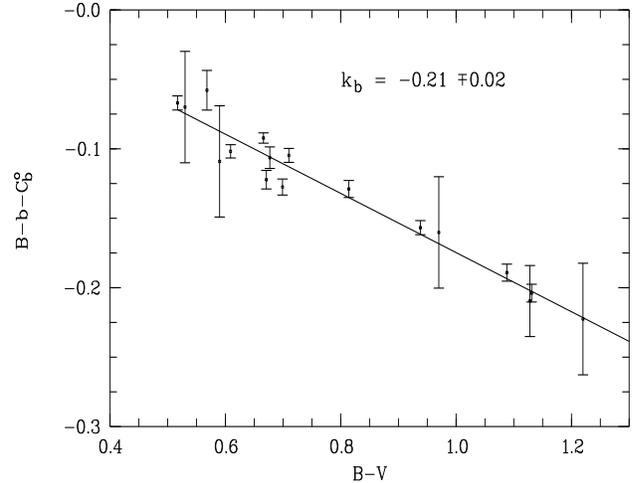,angle=-90,height=7.5cm,width=9.cm}}
       \caption[]{Colour term as a function of colour for standard star
sequences of an entire observing run (NTT/EMMI-B, CCD TEK\#31)
. The error bars for each 
star is the quadratic sum of the instrumental error resulting from several
measurements and the intrinsic error given by the authors. The solid line 
represents the weighted least-squares regression whose slope provides the 
value of $k_B$. The $C_b^0$ are chosen for each set of stars of each night to 
 minimize the scatter in the colour term $k_b$.}
            \label{km}
    \end{figure}

\begin{figure}
      
\centerline{\psfig{figure=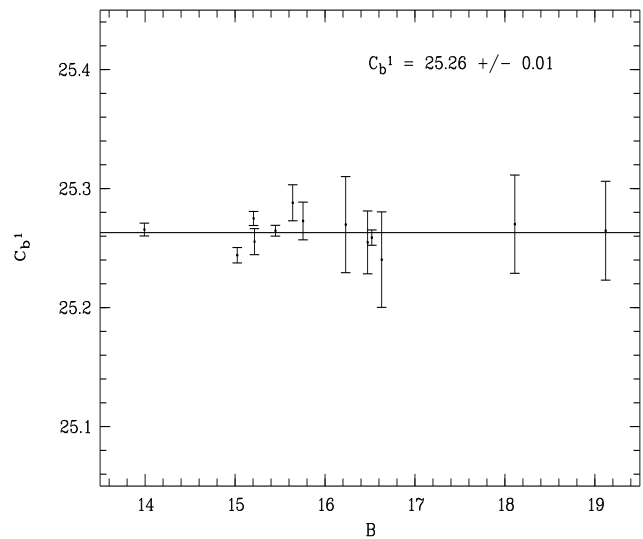,angle=-90,height=7.5cm,width=9.cm}}
       \caption[]{ Resulting  zero-point value $C_b^1$ for  one observing night 
 using the color coefficient $k_B$ calculated in  Fig.~\ref{km}. 
The error bars for stars are the same as in Fig.~\ref{km}, and the zero-point
 is calculated by weighted least-squares regression by a constant term.}
            \label{zp}
    \end{figure}

The zero-points $C_M$ are specific to each combination of
Telescope/Instrument/CCD/Filter/atmospheric conditions. These are
easily measured using a large number of standard stars observed each
night (see Fig.~\ref{zp}).  The colour coefficients $k_M$ allow the
magnitudes from the ``observing'' filter (resulting from the
Telescope/Instrument/CCD/Filter combination) to be corrected into the
B, V, R standard filters.  We assume that they remain constant during
each observing run, and use the calibrations of an entire run to
calculate these coefficients.  Two color coefficients for the V band
can be calculated depending on whether B$-$V or V$-$R colours are used
(denoted $k_{V_B}$ or $k_{V_R}$ respectively). \\ 
In practice, we
estimate the colour coefficients for each different configuration of
Telescope /Instrument /CCD /Filter as shown in Fig.~\ref{km}.  The
resulting colour coefficients are shown in Table 3. The quoted errors
are the r.m.s. uncertainties in the linear fit.  The measured colour
coefficients are then used to determine the accurate zero-point $C_M$
for each night. \\

%--------------------------------------------------------
%--------------------------------------------------------
%--------------------------------------------------------
%--------------------------------------------------------  
%--------------------------------------------------------
\begin{table*}
  \label{tabcolcoef}
  \caption{Measured colour coefficients}
  \begin{tabular}{lllll}
  \hline
    Inst/CCD \ \ \ & $k_B$  &  $k_{V_B}$ & $k_{V_R}$ & $k_R$ \\
  \hline      
    EFOSC/RCA\#8 \ &0.16$\pm$.03  & 0.04$\pm$.02& 0.10$\pm$.02 & 0.00$\pm$.02  \\
    EMMI-R/THX\#18  &              & 0.05$\pm$.01& 0.10$\pm$.02& -0.10$\pm$.01 \\
    EMMI-R/LOR\#34  &              &             &              & -0.03$\pm$.01 \\
    EMMI-R/TEK\#36  &              & 0.03$\pm$.02& 0.05$\pm$.02&                \\
    EMMI-B/TEK\#31  &-0.21$\pm$.02 &             &             &                \\
  \hline
  \end{tabular}
\end{table*}
%--------------------------------------------------------

\section{STANDARD MAGNITUDES}
\subsection{Transformation equations}
 The calibration of the five different CCDs used in the imaging survey
allows us to derive a homogeneous set of standard magnitudes for all
detected objects.  Instrumental magnitudes are related to standard
magnitudes by the transformation equations obtained from Eq.1 :
%--------------------------------------------------------
\begin{equation}
 M_{std} = M_{ccd} + k_M (B-V) 
\end{equation}
%--------------------------------------------------------

where $M_{ccd} = m_{ccd} - A_M sec\xi + C_M$ \\ 

We assume that no correction for galactic absorption is necessary
because our fields lie at high galactic latitude $b^{II} \sim
-83^{\circ}$ and the widest extent of the full survey is only 1.53
deg. \\

Because we have a priori no knowledge of the colours of the detected
objects in the Johnson-Cousins system, we invert the set of equations
(2) so that the colour correction appears in terms of {\it observed}
colour. We give the equations for the V band where two colour
coefficients are measured :
%--------------------------------------------------------
\begin{equation}
 V = V_{ccd} + \frac{k_{V_B}}{1 - k_B + k_{V_B}}.(B_{ccd} - V_{ccd}) 
\end{equation}
\begin{equation}
 V = V_{ccd} + \frac{k_{V_R}}{1 - k_{V_R} + k_R}.(V_{ccd} - R_{ccd}) 
\end{equation}
%--------------------------------------------------------

Most objects detected in one filter are identified in the other 2
filters (see next subsection).  For the galaxies detected in only one
filter, we derive the magnitudes in the other filters using the
standard colours B$-$V = 1.0 and V$-$R = 0.5.  These mean values are
derived from the mean galaxy colours obtained for the survey and
reported in section 9.2.2.  To check the reliability of our colour
coefficients listed in Table 3 and the resulting standard magnitudes,
we estimate V in two possible ways using equations (3) and (4). The
mean differences in the 2 corresponding magnitudes for 7000 objects
observed at the NTT and 5000 objects at the 3m60 are $\le$ 0.005 $\pm$
0.001 mag.  This test demonstrates the reliability of our measured
colour coefficients.  Because the R and V frames cover almost the same
area of the sky, the final V magnitude are calculated in terms of
V$-$R colour (Equation 4).
\subsection{Colour completeness}
\begin{figure*}     
\centerline{\psfig{figure=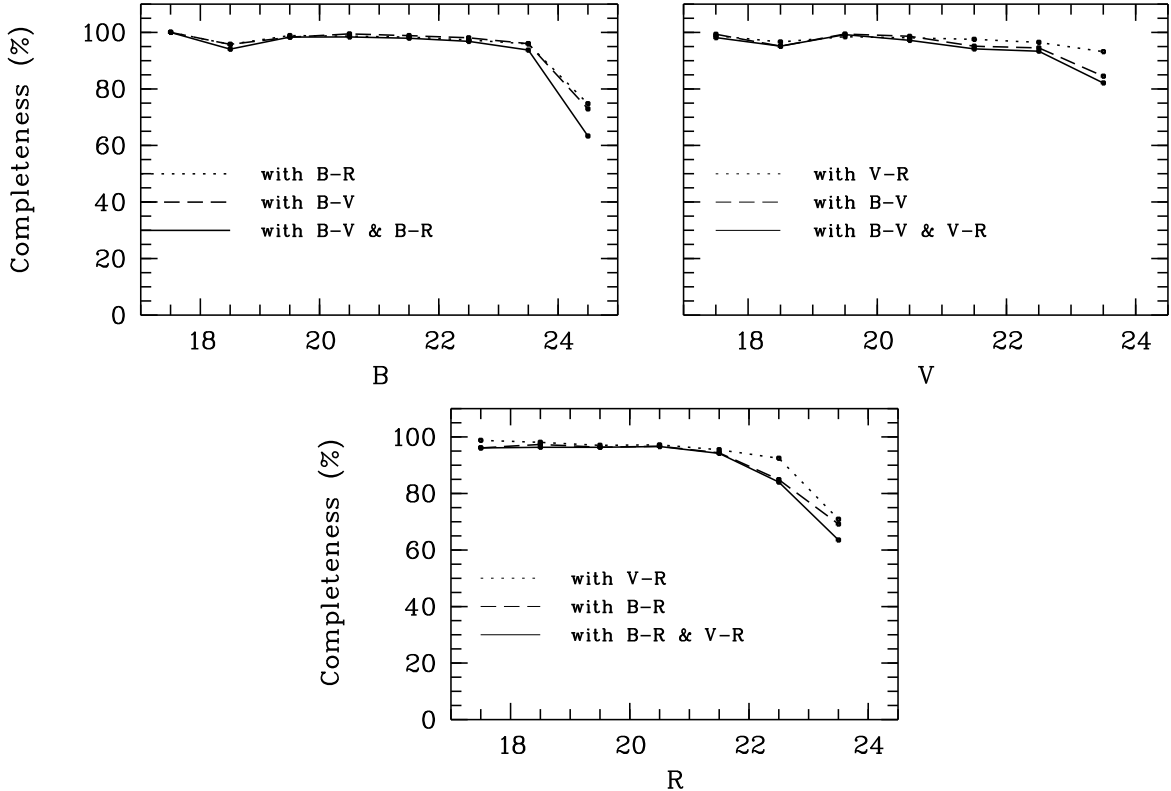,angle=-90,height=12cm}}
       \caption[]{ Colour completeness of B, V and R detected objects 
 in B$-$V and V$-$R  colours. The completeness is estimated in  bin sizes of 
 1 mag. The coding of the curves is specified inside the graph.}
            \label{complB}
    \end{figure*}

Although R and V frames cover almost the same area of the sky, the B
frames cover only about 80\% of area observed in R and V, partly
because of the smaller CCD size in the EMMI blue channel.  Within the
common area of the B, V and R bands, we define a colour completeness
rate for objects selected in the B band as the fraction of B detected
objects with associated V or R or both detections (in bins of 1
mag). We calculate the analogous curves for the V and R bands inside
the same area. These various curves are shown in Fig.~\ref{complB}.
Typically, the completeness drops below 90\% 1 mag brighter than the
limit of the catalogue in each band. At the limit of the spectroscopic
catalogue R$\le$20.5, the completeness in B and V is greater than
95\%.
%
%--------------------------------------------------------
%--------------------------------------------------------
\section{MATCHING CCD OVERLAPS}

Because our images were taken over a 6-year period using different
telescopes (3.60m and NTT), instruments (EFOSC2 and EMMI) and CCDs (
different pixel size and color coefficients), it is crucial to adjust
the zero-points of our magnitude scales over the whole survey. This is
done by comparing the measured magnitudes of objects located in the
overlapping edges of the CCD images (1/10 of the CCD size, i.e. $\le$
1 arcmin). This adjustment guarantees an internally homogeneous
photometry, and corrects for the systematic offsets in the photometric
calibrations from one observing night/run to another, as well as other
possible variations not accounted for like the systematic variations
in the extinction curves due to the eruption of the Pinatubo volcano
in 1991 (Burki et al. 1995).\\ 
The input images have the standard
magnitudes derived in section 7, and the equatorial coordinates
($\alpha$, $\delta$) described in section 5.  Because the overlaps
contain few objects, we use all stars and galaxies with R $<$ 22 mag.
The magnitudes can then be directly compared in the overlaps of each
CCD, and we show below how the mean offsets provide the correction of
zero-point to apply to each CCD.
%--------------------------------------------------------
%      
\subsection{Method}
Because each CCD frame has 2 to 4 overlaps, the adjustment of
zero-points must be done by a global least-squares fit.  We use the
method described by Maddox et al. (1990a) and originally proposed by
Seldner et al. (1977).  Seldner designed this method for adjustment of
galaxy counts in the overlapping cells on neighbouring plates to
correct the systematic variations in the magnitude limits from plate
to plate. Instead of comparing the counts, Maddox formalism allows to
compare the differences of magnitude for common objects on overlapping
photographic plates.  For a complete description of the technique the
reader can refer to Maddox et al. (1990a).  Here we briefly review the
principle of the method. \\ 

\begin{figure}     
\centerline{\psfig{figure=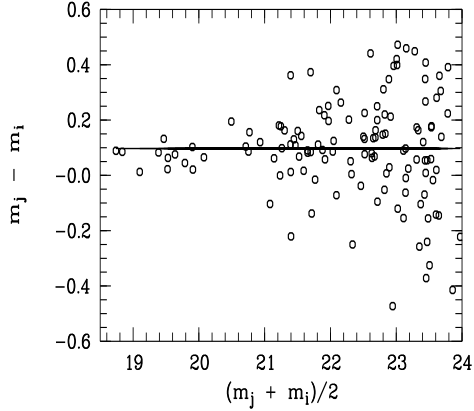,angle=-90,height=8.cm,width=8.8cm}}
       \caption[]{Difference of magnitudes for common objects in two overlapping
  CCDs j and i. The solid line shows the linear fit for objects brighter than
 22 mag which measures $\overline T_{ij}= m_j-m_i$.}
            \label{meanof}
    \end{figure}

For each galaxy k on CCD i, we define a correction for the zero point
$\delta C_i$ by

\begin{equation}
m^k_0 = m^k_i + \varepsilon_i + \delta C_i
\end{equation}

where $m^k_0$ is the ``true magnitude'' and $m^k_i$ is the measured
magnitude with an error $\varepsilon_i$.  On each overlapping CCD pair
(i,j), we obtain
\begin{equation}
m^k_0 = m^k_i + \varepsilon_i + \delta C_i = m^k_j + \varepsilon_j + \delta C_j
\end{equation} 
 We define a mean offset $\overline T_{ij}$ so that
\begin{equation}
\overline T_{ij} = <m^k_j - m^k_i> \simeq \delta C_i - \delta C_j
\end{equation}

here, we assume that $<\varepsilon_i>=<\varepsilon_j> = 0$.  Because
the CCD provides a linear response in a large magnitude range,
$\overline T_{ij}$ is independent of the magnitude ($m_j + m_i$)/2.
We thus measure $\overline T_{ij}$ by an offset in the zero$^{th}$
order of ($m_j - m_i$) versus ($m_j + m_i$)/2 (for the photographic
plates, Maddox \& al (1990a) define $\overline T_{ij}$ by a third
order adjustment).  Fig.~\ref{meanof} shows an example of the
differences of magnitudes for all objects in one overlap (i,j).  We
require that the overlaps taken into account in the equation system
have more than 5 objects in order to guarantee a reliable measure of
the mean offset value.\\ 
For CCD i, we obtain several correction
factors corresponding to each of its neighbours j which we denote as
\begin{equation}
 \delta C_i = \delta C_{i_j} = \delta C_j + \overline T_{ij} \quad \quad \quad \quad \quad \quad j \subset i
\end{equation}

where $\delta C_{i_j}$ is the correction factor estimated with
neighbouring CCD j, and $j \subset i$ denotes all the neighbours j of
CCD i.

The goal is to determine the unique value $\delta C_i$ for CCD i which
globally minimizes the scatter in the $\delta C_{i_j}$. These
coefficients are calculated by minimization of the function F defined
by:
\begin{equation}
F = \frac{1}{2} \sum_i \sum_{j \subset i} W_{ij} ( \overline T_{ij} +
\delta C_j - \delta C_i)^2 
\end{equation}

where $W_{ij} = 1/var(T_{ij})$ are the weighting factors which favor
the overlaps with a small variance. \\ 
With $\partial F \over
{\partial \delta C_i}$ = 0, we obtain the following set of equations :
\begin{equation}
\delta C^{n+1}_i = \frac{ \displaystyle\sum_{j \subset i}( \delta C^n_j +
\overline T_{ij}) W_{ij}}{\displaystyle\sum_{j \subset i} W_{ij}} 
\end{equation}

which are solved by iteration.  To increase the convergence speed and
the stability of this iterative method, we adopt the technique from
Maddox of adding the previous value $\delta C_i^n$ into equation (10):
\begin{equation}
\delta C^{n+1}_i = \frac{\left ( \delta C^n_i W_{ii} + \displaystyle\sum_{j 
\subset i}( \delta C^n_j + \overline T_{ij})
W_{ij} \right )}{\left (W_{ii} + \displaystyle\sum_{j \subset i} W_{ij} \right )} 
\end{equation}
where $W_{ii}$ is the mean value of all $W_{ij}$ of each CCD i.
%--------------------------------------------------------
%------
%--------------------------------------------------------
    \begin{figure}      
\centerline{\psfig{figure=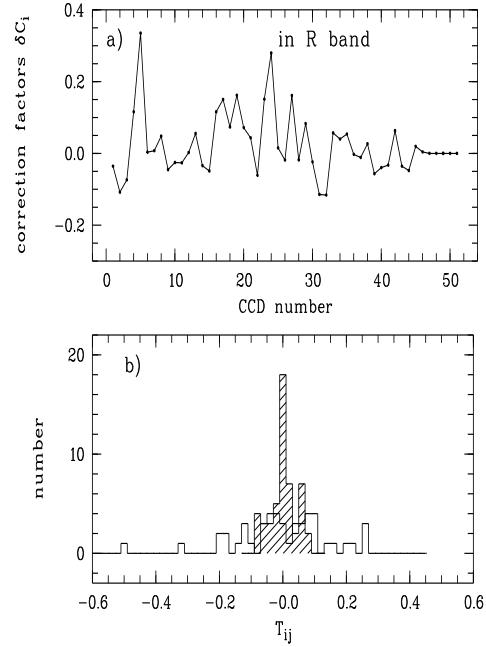,angle=-90,height=9cm,width=9.2cm}}
       \caption[]{The upper graph shows the estimated magnitude correction 
 for the R filter to be applied to each CCD zero-point. The  first 30 
CCDs are fields observed at the 3m60 and the remaining come from NTT observations.
  CCDs from 47 to 50 are  used as reference frames with no correction factors which
 provide the constraints to the other frames
  The lower graph shows the histogram of $\overline T_{ij}$ before correction of
 the zero-point  (empty histogram) and after correction (hashed histogram). }
            \label{CipR}
    \end{figure}
%--------------------------------------------------------
\subsection{Results}

To constrain the set of correction factors, Maddox used $\sum \delta
C_i = 0$. In our case, we have re-observed several fields across the
whole survey during the last observing run: 4 fields in B with
TEK\#31, 4 in V with TEK\#36 and 5 fields in R with LOR\#34.  The
seeing conditions were good ($FWHM \sim 1.0 \arcsec$, $1.0\arcsec$,
$0.8\arcsec$ in B, V, R respectively) and the photometric calibrations
were performed with great caution.  We use these fields as fixed
references for the adjustments of the zero-points of the other
frames. During this process, we detected systematic offsets in the
initial V and R zero-points of the 3.6m frames derived from
calibration sequences (section 6) when compared with the NTT reference
frames: $\delta C_{3m60} - \delta C_{NTT} = +0.10^m$ in the R band,
$\delta C_{3m60} - \delta C_{NTT} = -0.10^m$ in the V band.  These
shifts cannot be explain by a problem in the colour correction because
the offsets are independent of the colour of the objects.  Because of
the absence of a similar shift in the B band, this effect cannot be
attributed to the variation of the extinction coefficients due to the
eruption of the Pinatubo volcano in 1991, which occurred when the
observations switched from the 3m60 telescope to the NTT telescope.
To prevent the minimization method from introducing a gradient in the
set of correction factors around the boundary between NTT and 3m60
frames the equation systems in each band for the NTT and 3m60 are
processed separately.  \\ 
First, we use $\sum \delta C_i = 0$ as a
constraint for both systems. The solutions of the two systems provide
the internal zero-point corrections to be applied to each CCD.  Then,
we use the reference fields to derive the global magnitude shift of
each ensemble of zero-points (3m60 and NTT). These various steps allow
us to adjust the two systems to the same zero-point.  The resulting
correction factors for the R band are shown in Fig.~\ref{CipR} a).
Each point defines one CCD and corresponds (with some gaps) to a
sequence number (see figure captions).  The iteration of Eq. (11) is
stopped when the difference between the previous and new estimation
($\delta C_i^{n+1}- \delta C^n_i$) is smaller than 0.005$^m$ for all
CCD i.  The typical number of required iterations is 10. \\ 
In
Fig.~\ref{CipR} b), we plot the histogram of initial and corrected
$\overline T_{ij}$, $\overline T_{ij}^{cor}$ defined as :
\begin{equation} \overline T_{ij}^{cor} = \overline T_{ij} + \delta
C_j - \delta C_i \end{equation} In these histograms, we exclude the
symmetric terms $\overline T_{ji} = - \overline T_{ij}$. \\ 
The r.m.s scatter in the initial $\overline T_{ij}$ in B, V, R bands
are respectively 0.11$^m$, 0.13$^m$, 0.14$^m$. After correction by
$\delta C_i$, we reduce the scatter to 0.04$^m$ in all 3 bands.
\subsection{Conclusion}

These results demonstrate the efficiency of the method to reduce the
dispersion between the CCD frames over the whole survey. They also
allow us to estimate the dispersion in our magnitude system.  In fact,
if there was no error in the measures of magnitudes, $\overline
T_{ij}$ would be equal to zero.  The dispersion in the $\overline
T_{ij}$ after adjustments gives a 0.04$^m$ error in our measured
magnitudes in all 3 bands and is in good agreement with the 0.05$^m$
in the adaptive aperture magnitude uncertainty given by the
simulations (see section 4.2).  To check the consistency of zero-point
calibrations between the NTT and 3m60 fields, we have applied a K-S
two-sample test to the B$-$R and B$-$V colour distributions obtained
as described in last section.  The probability that the 2
distributions are drawn from the same parent distribution is 0.45
after correction of the systematic zero-point offsets between 3m60 and
NTT frames, it is 0.01 when the correction is not applied. This
confirms that correction of the offsets in zero-point for the 3m60 in
R and V bands is necessary. Finally, there could be a systematic shift
in our overall zero-point, but it would be within the error bars
estimated from our individual zero-points (i.e. $\le$ 0.05$^m$).
 
\section{FIRST RESULTS}
\subsection{Star sample}
\begin{figure*}     
\centerline{\psfig{figure=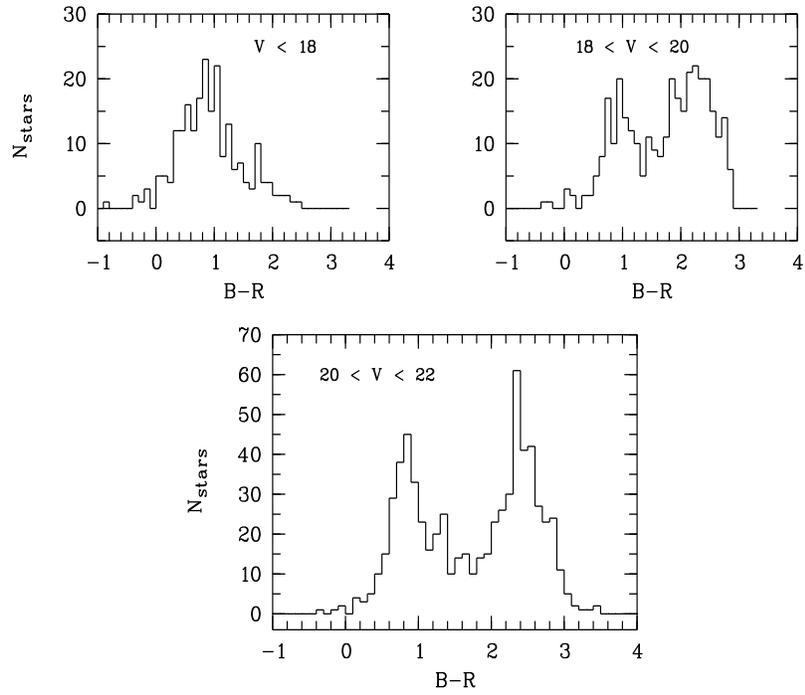,angle=-90,height=10cm,width=13.cm}}
       \caption[]{ B$-$R colour histogram for the stars selected in three different
 ranges of V magnitude as specified within the graphs.}
            \label{brstar}
    \end{figure*}

\begin{figure*}     
\centerline{\psfig{figure=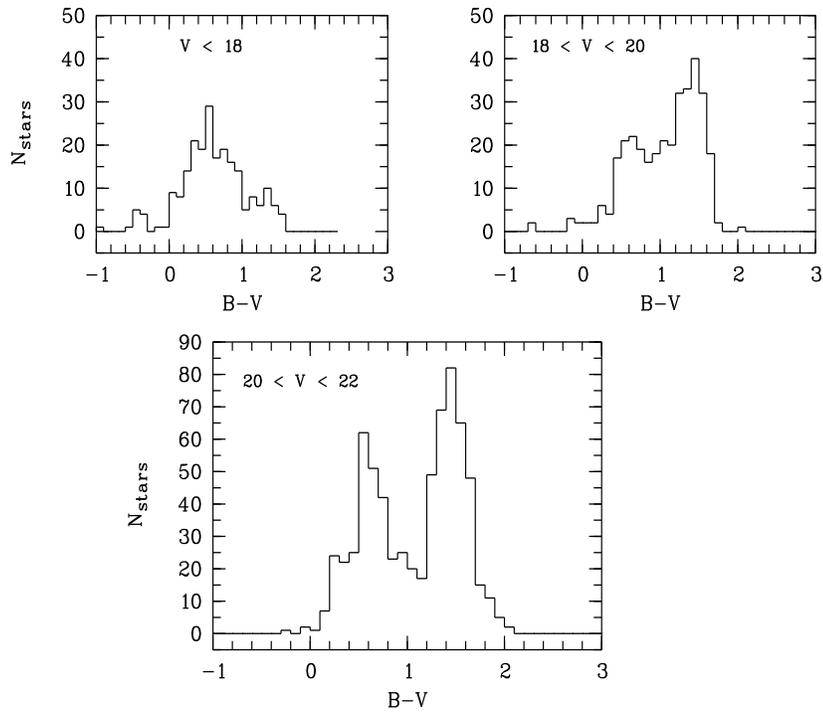,angle=-90,height=10cm,width=13.cm}}
       \caption[]{same as Fig.~\ref{brstar} for B$-$V colour.}
            \label{bvstar}
    \end{figure*}

Byproducts of deep imaging surveys at different galactic latitudes are
stellar samples in different directions of our galaxy at very faint
magnitudes (Shanks et al. 1980, Kron 1980, Infante 1986, Metcalfe et
al. 1991).  In Figures ~\ref{brstar},~\ref{bvstar}, we show for our
stellar sample the different colour histograms B$-$R, B$-$V, for three
magnitude ranges in V. Note that the quasars are included in this
sample. Also, at bright magnitudes (V $<$ 17), a large fraction of
stars are saturated, thus raising the uncertainty in the corresponding
colours. \\ 
At faint magnitudes ($V \ge 18$), two stellar populations
can be seen.  Brightward of $V < 18$, one broad blue peak is present
near B$-$V$\sim$0.5-0.6.  Faintward of this magnitude a second peak
appears in the red part near B$-$V$\sim$ 1.4-1.5.  These results agree
well with the other existing data (see table 4 above).  The red peak
is interpreted as being nearby M stars belonging to the disk
population.  The blue sequence is interpreted as being stars belonging
to the galactic halo (Robin et Cr\'ez\'e 1986).\\ 
These observations
are listed in Table 4 for $V\le$ 22. The second column gives the
colours used by the author, the third and fourth columns show the
colours of the two peaks in this initial colour and the subsequent
columns show the transformed position of the two peaks into the
standard system Johnson-Cousins.  The transformation equations are
given in the corresponding publications, and for the Kron (1980) data
we added the transformation equations given by Majewski (1992) as
(J$-$F)=0.738(B$-$R) -0.02.  Table 4 shows that our stellar color
distributions are in good agreement with those resulting from other
faint CCD surveys.
%--------------------------------------------------------
\begin{table*}
  \label{tabcolstar}
  \caption{stellar colour peak for V$\le$22}
  \begin{tabular}{llccccccccc}
  \hline
    Reference  & Observed & \multicolumn{2} {c} {blue peak  red peak}  &  
 \multicolumn{3} {c} {blue peak}  & \multicolumn{3} {c} {red peak} \\
               &  colours &   &  & B$-$V & B$-$R & V$-$R &  B$-$V & B$-$R & V$-$R  \\
  \hline
 Kron et al. (1980)  &J$-$F & 0.65 & 1.75  & 0.50 & 0.90 & 0.40 & 1.50 & 2.40 & 0.90\\
 Shanks et al. (1980)&J$-$R & 1.0 & 2.2 &  0.6   & &  & 1.6 & & \\
 Infante (1986)      &J$-$r & 0.9 & 2.3 & 0.6 & &  & 1.5 & &  \\
 Metcalfe et al. (1991)&$(B$-$R)_{ccd}$ & 1.0 & 2.2 & & 1.12 &  & &2.46 &  \\
 This work           &  &  & & 0.55 & 0.90 & 0.35 & 1.45 & 2.40 & 0.95 \\   
  \hline
  \end{tabular}
\end{table*}
%--------------------------------------------------------

%--------------------------------------------------------
%--------------------------------------------------------
%--------------------------------------------------------
%--------------------------------------------------------
\subsection{Galaxy sample}
\subsubsection{Galaxy counts}
\begin{figure*}     
  \centering  
  \hbox{
  \subfigure[Galaxy counts in B band]{ 
  \psfig{figure=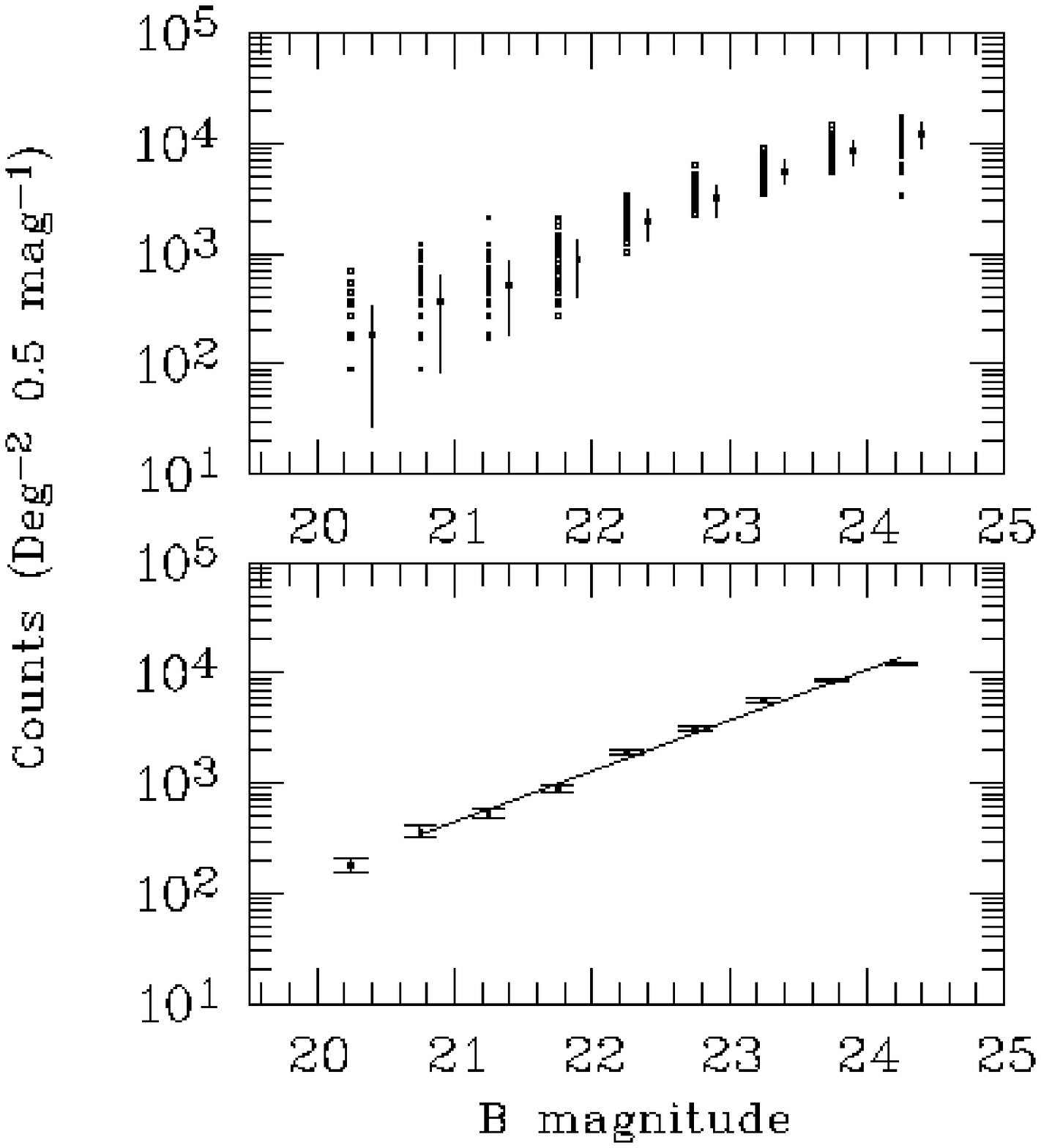,height=9.cm,width=6.cm,angle=-90}}
  \subfigure[Galaxy counts in V band]{ 
   \psfig{figure=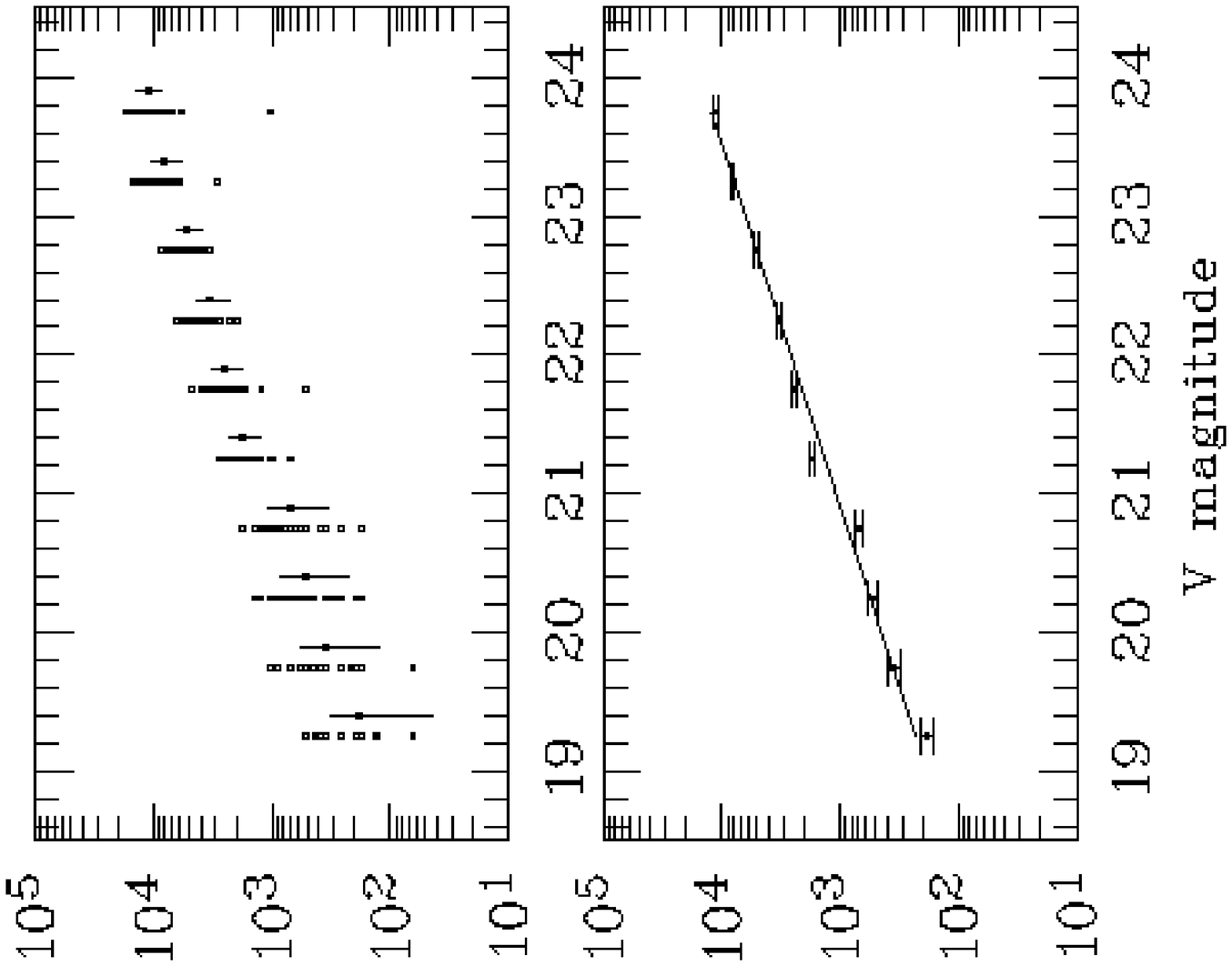,height=9.cm,width=5.5cm,angle=-90}}
  \subfigure[Galaxy counts in R band]{ 
   \psfig{figure=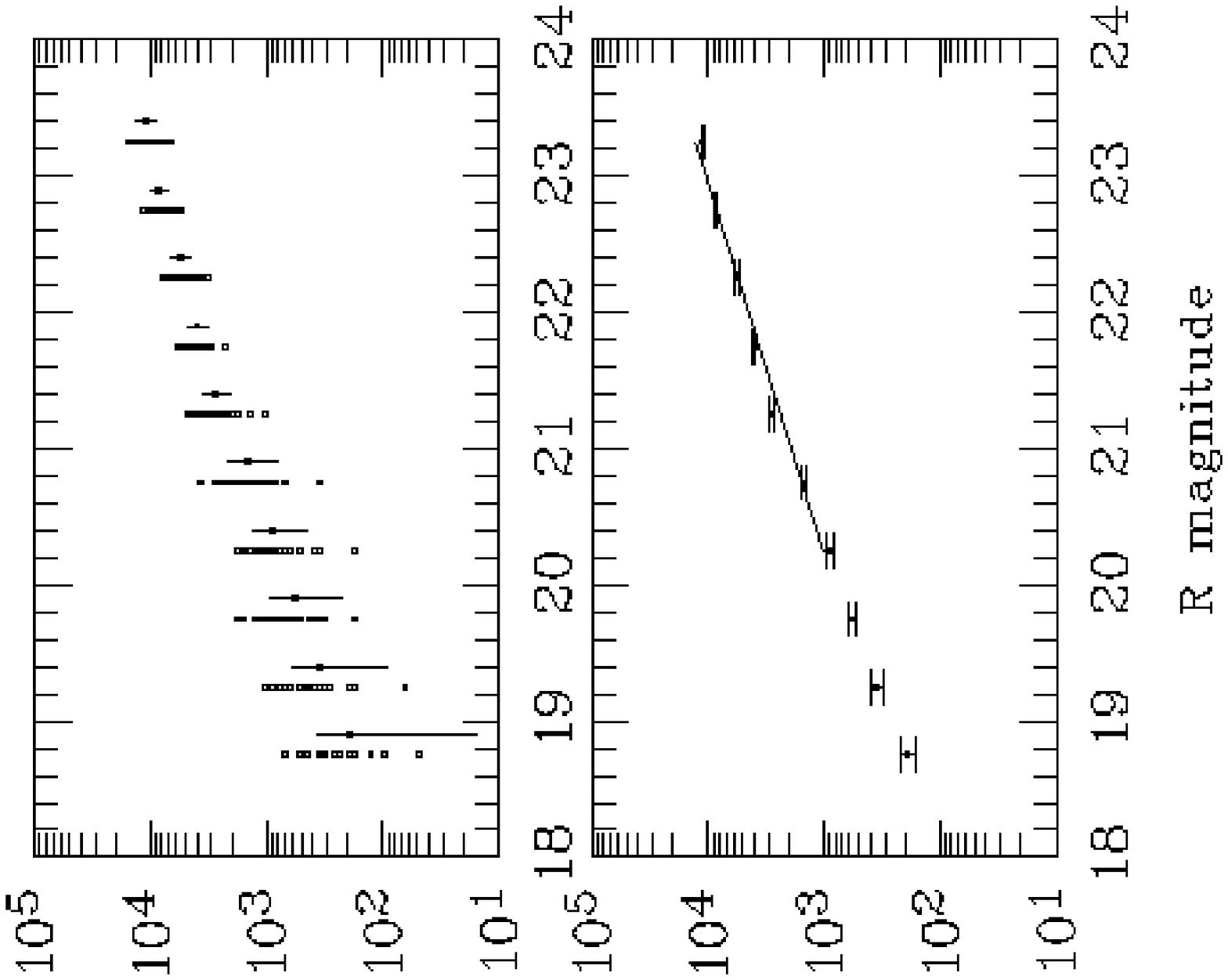,height=9cm,width=5.5cm,angle=-90}}
        }
       \caption{Galaxy number-magnitude counts by square degree per 0.5 mag 
interval in the B (left), V(center) and R (right) bands.
 The upper graph shows the counts for each observed CCD field and the median values
with error bars given as the $\sigma$ field-to-field fluctuation. 
The lower graph shows the median number count value in each bin. 
The error bars are given by equation (13). 
 The solid line is the estimated slope  by least squares fit.}
    \label{bgal}
    \end{figure*}

 a) Counts and slopes \\
We present the differential galaxy counts based on the total 0.4
square degree of the survey in the R, V and B bands to limiting
magnitudes of 23.5, 24.0 and 24.5 respectively. These limiting
magnitudes are defined as the last magnitude bin ($\Delta$m=0.5 mag)
before the turn-off in the number counts.  To these limits the galaxy
catalogs contain about 13000, 12150 and 9500 objects in the R, V and B
bands respectively.  Note that there is no star-galaxy separation for
R $\ge$ 22 and no correction for stellar contamination is done because
the expected number of stars at this galactic latitude ($b^{II}\simeq
-83^{\circ}$)is lower than 4\% (corresponding to an offset of $\sim$
0.017 of the logarithmic number counts).\\ 
In Fig.~\ref{bgal}, we show the differential galaxy number counts per
square degree in 0.5 magnitude bins.  The upper graphs show for each
band the superimposed galaxy counts for all individual CCD fields. The
dots displaced to the right of the columns of galaxy counts are the
median counts offset by 0.2 mag for clarity.  The error bars measure
the 1$\sigma$ rms field-to-field scatter.  The field-to-field scatter
is $\sim$ 20 \% in the R band between 21 $<$ R $<$ 23.5, $\sim$ 25 \%
in the V band between 21 $<$ V $<$ 24. and $\sim$ 30 \% in the B band
between 22 $<$ B $<$ 24.5.  These r.m.s. dispersions significantly
exceed the expected Poisson variations in $\sqrt{N}$ because of
galaxy-galaxy clustering (Arnouts \& de Lapparent 1996).  In the
bright part, the large values of the scatter are essentially due to
the small number of bright galaxies per CCD frame.  The lower graphs
give the median number counts.  In these graphs the error bars are
given as:
\begin{equation} 
\sigma = \sigma_{field-to-field} / \sqrt{N_{field}}
\end{equation}

\begin{figure*}
  \centering  
  \hbox{
  \subfigure[Comparison with others deep CCD galaxy number counts]{ 
  \psfig{figure=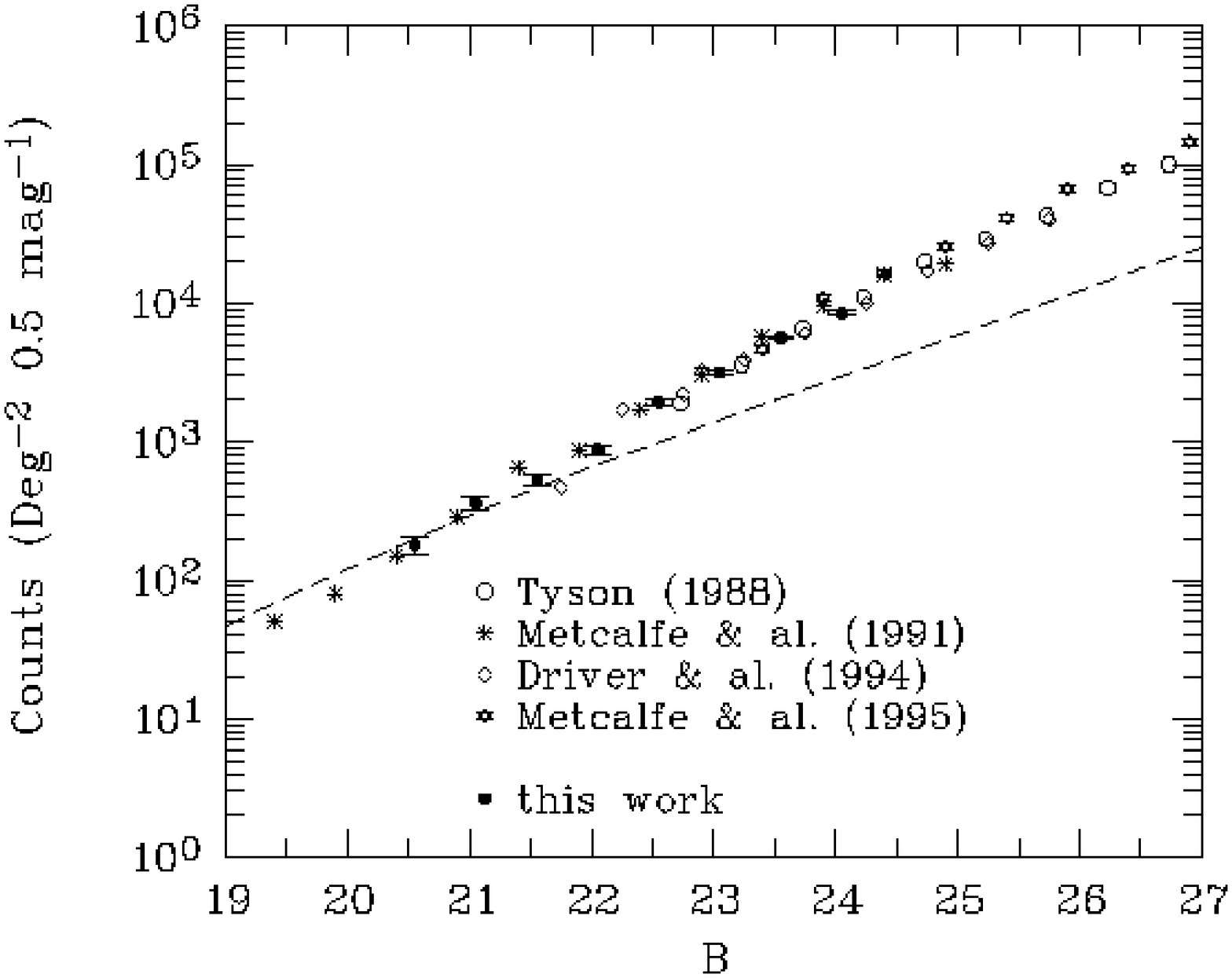,height=7.cm,angle=-90}}
  \subfigure[Comparison with bright photographic galaxy number counts from 
Bertin et Dennefeld (1996)]{ 
   \psfig{figure=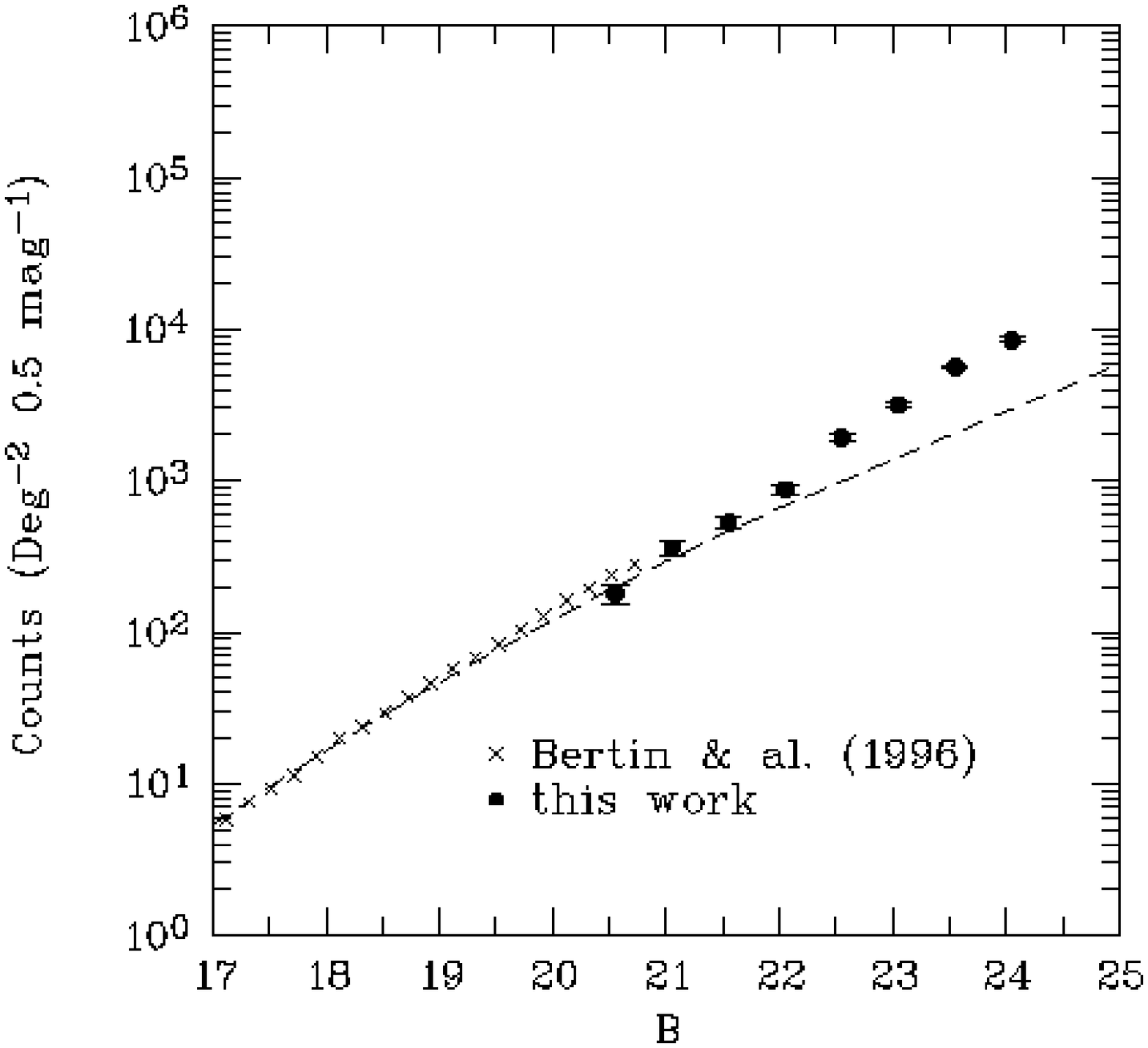,height=7.5cm,angle=-90}}
        }
  \caption{Comparison with others published  galaxy number 
counts transformed into the  B Johnson filter. The error bars for our data show 
the  $\sigma$ estimated in Eq. 13. The dashed line shows the differential number 
counts  expected for a non evolving model using a $\Phi^{\star}$ normalized to the
bright galaxy number counts from Bertin et Dennefeld (1996) as described in the 
 text.}
  \label{bjgal}
\end{figure*}

\begin{figure*}     
\centerline{\psfig{figure=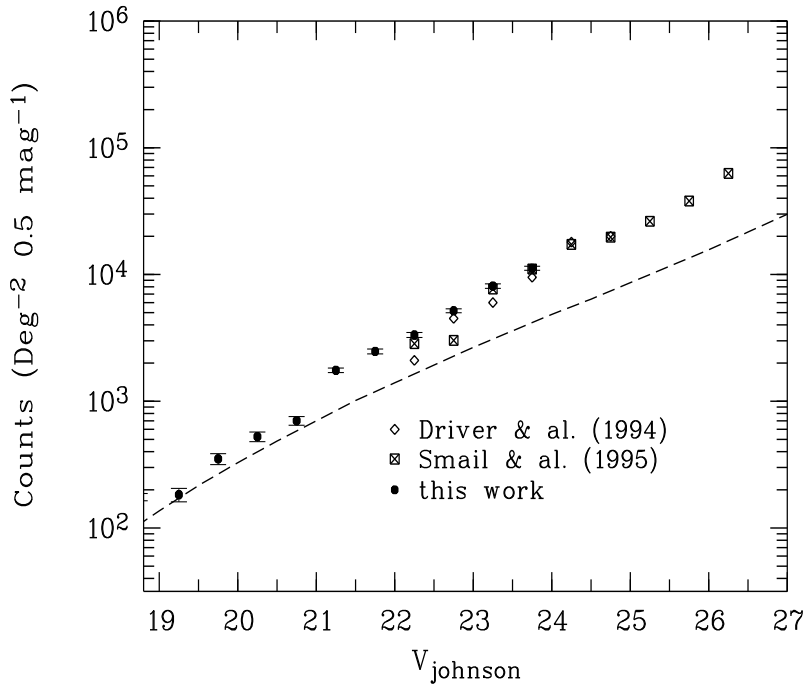,angle=-90,height=11.cm,width=14.cm}}
       \caption[]{ Same as Fig.~\ref{bjgal} in the V Johnson filter.}
            \label{vjgal}
    \end{figure*}

\begin{figure*}
  \centering  
  \hbox{
  \subfigure[Comparison with others deep CCD galaxy number counts]{ 
  \psfig{figure=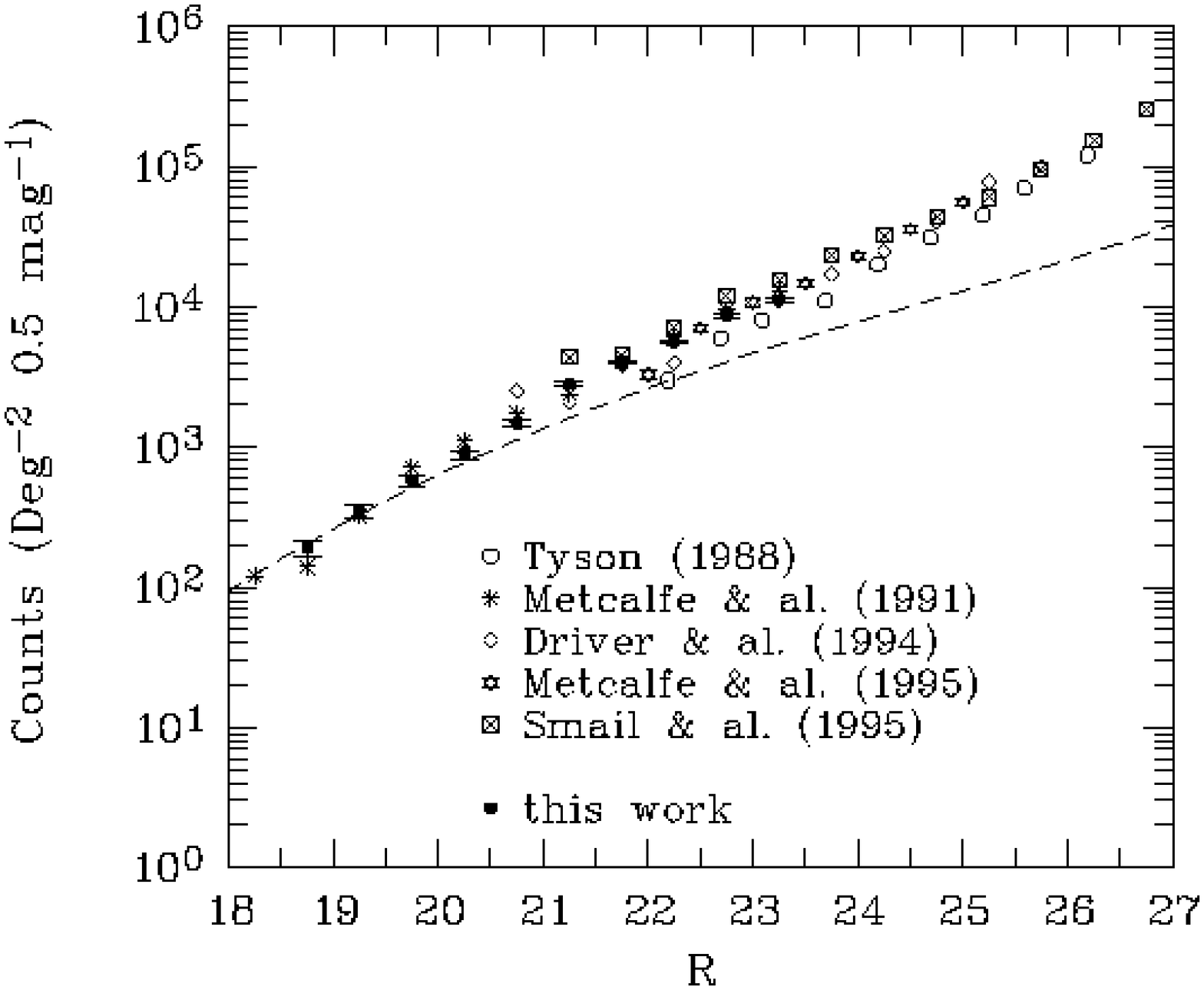,height=7.5cm,angle=-90}}
  \subfigure[Comparison with bright photographic galaxy number counts from Bertin
 et Dennefeld (1996)]{ 
   \psfig{figure=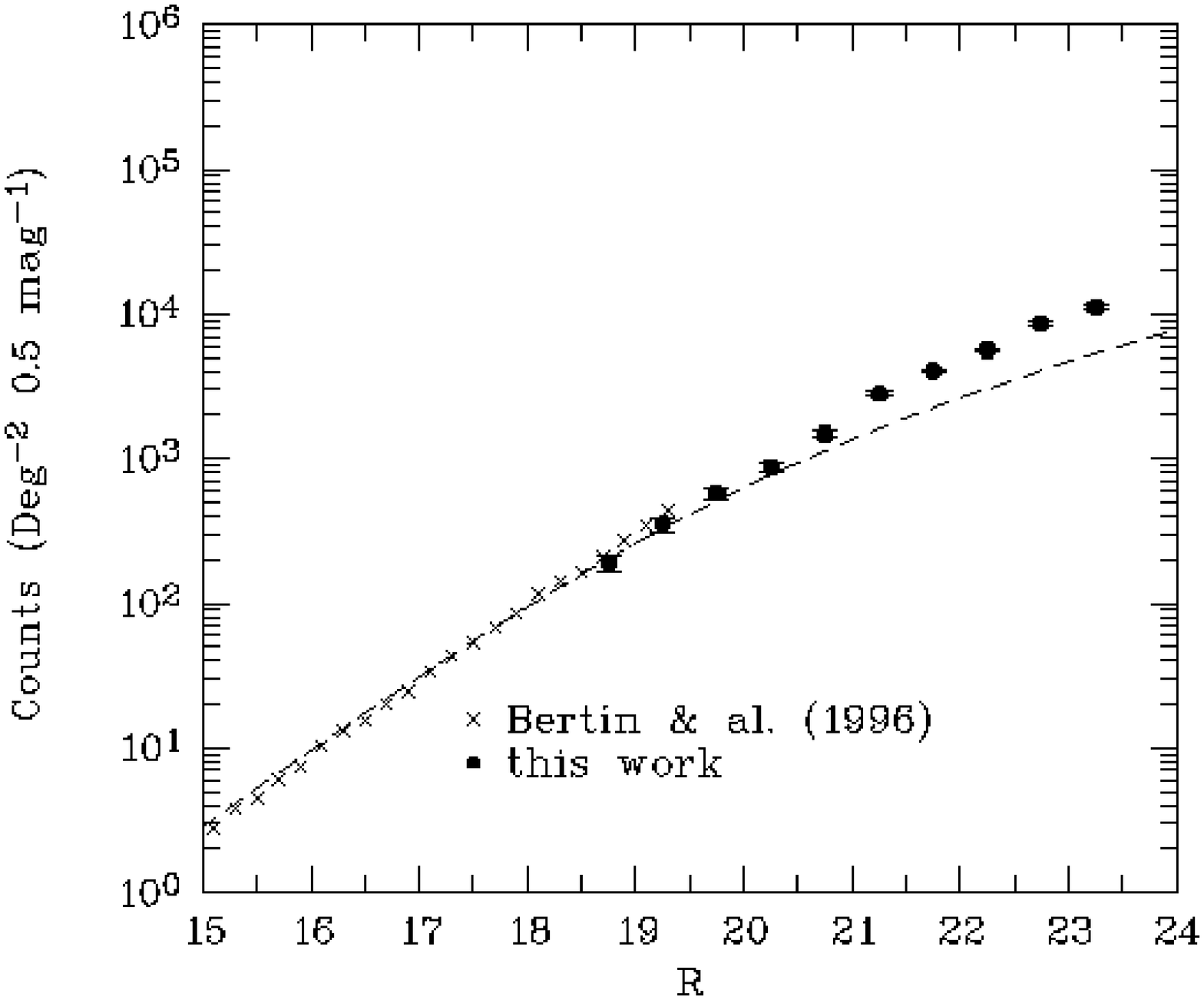,height=7.5cm,angle=-90}}
        }
  \caption{ Same as Fig.~\ref{bjgal} in the R Cousins filter}
  \label{rfgal}
\end{figure*}

The differential number counts (in deg$^{-2}$ 0.5 mag$^{-1}$) are
fitted by a power law in the same magnitude range as Metcalfe et
al. (1991) for the R and B magnitudes.  The exponent of the fit in the
three bands are measured by a least squares fit in the logN-mag plots.
These fits are shown by solid lines in the lower graphs of
Fig.~\ref{bgal} and are parameterized as follow :
\begin{equation}
 N_{gal}= 10^{(0.460 \pm 0.02) B} \times 10^{-7.01 \pm 0.39}
\end{equation} 
for 20.5 $\leq B \leq $ 24.5
\begin{equation}
N_{gal}= 10^{(0.384 \pm 0.02) V} \times 10^{-5.03 \pm 0.47} 
\end{equation}
for 20.0 $\leq V \leq $ 24.0
\begin{equation}
N_{gal}= 10^{(0.367 \pm 0.02) R} \times 10^{-4.42 \pm 0.47} 
\end{equation}
for 20.0 $\leq R \leq$ 23.5 \\
In table 5, we summarize the results of previous CCD surveys on galaxy
counts in the visible bands.  Our slopes are in good agreement with
the other works. \\ 
In Fig 12, we detect two magnitude bins between
21.0 $\le B \le$ 22.0 where the density systematically decreases and
this effect is seen in the three bands (in the intervals 20.5 $\le V
\le$ 21.5 and 20.5 $\le R \le$ 21.5). First of all, to see if this gap
was caused by inhomogeneities in the projected distributions, we
examined the angular distribution (RA, Dec) in different magnitude
intervals, but no particular feature in the clustering of the
projected distributions was visually detected.  This investigation
will be pursued quantitatively in a forthcoming study of the two-point
angular correlation function for these data (Arnouts \& de Lapparent
1996).\\
%--------------------------------------------------------
%--------------------------------------------------------
%--------------------------------------------------------
\begin{table*}
  \label{tabslope}
  \caption{The galaxy number count exponents for several CCD photometric surveys}
  \begin{tabular}{lccc}
  \hline
  Reference \ \ \ & B  &  V  & R \\
  \hline
  Tyson (1988)$^{a}$   &  0.45 &  & 0.39 \\
  Lilly et al. (1988)$^{b}$  &  0.38 &  &  \\
  Metcalfe et al. (1991)$^{c}$ &  0.49 &  & 0.37  \\
  Driver et al. (1995)$^{d}$   &  0.44 & 0.40 & 0.37  \\
  Smail et al. (1995)$^e$          &  & 0.40 & 0.32 \\
 This work$^{e}$ &  0.46 & 0.38  & 0.37 \\   
  \hline
  a: in the $B_J$,R,I CCD photometric system & & & \\
  b: in the AB photometric system & & & \\
  c: in their CCD photometric system & & & \\
  d: in the KPNO system & & & \\
  e: in the Johnson-Cousins system & & & \\
  \end{tabular}
\end{table*}
%--------------------------------------------------------
 b) Counts comparison\\ 
In Fig.~\ref{bjgal},~\ref{vjgal},~\ref{rfgal}
we compare the differential number counts from our data with those
from the other CCD surveys.  For the data given in other systems than
the standard Johnson-Cousins system, we apply the different
transformations provided by the authors.  For the data from Metcalfe
we apply a transformation only for the B band as $B = B_{ccd} +
0.15$. \\ 
For the data from Tyson, the transformation equations are
not given. By default we use the transformations into the photographic
system ($b_J, r_F$) given in Metcalfe et al. (1991) combined with the
transformations from photographic bands to the standard system given
by Shanks et al. (1984). We obtain the approximate transformations
defined as $B = B_{Tyson} + 0.27$ and $R = R_{Tyson} - 0.07$.  \\ 
For
the data from Driver no transformations are done because the color
terms are small.  Except for Metcalfe et al. (1991), these other works
are significantly deeper than our data, and the number counts at very
faint magnitudes (B $\ge$ 25 and R $\ge$ 23.5) are corrected for
confusion.\\ 
Figures ~\ref{bjgal},~\ref{vjgal},~\ref{rfgal} show that
our counts in B and R are in good agreement with the results from
Metcalfe et al. (1991) in both the slope and the amplitude of the
logarithmic number counts in the red band but a small shift in the
blue band exists as we will see below in the discussion of the colour
distributions (section 9.2.2).  The R number counts of our survey are
significantly higher ($> 3\sigma$) than those from Tyson in the common
range of magnitudes.  This difference has been interpreted by Tyson
(1988) as being due to the a priori choice of fields devoid of bright
galaxies, and Metcalfe et al. (1991) suggest that this difference can
originate from the use of isophotal magnitudes by Tyson (1988) in
contrast to the ``total'' magnitudes used by the others authors.  The
data of Driver et al. (1994) also show a small deficit in galaxy
number counts compared to ours at the $\sim 3 \sigma$ level. Driver
specifies that the Hitchhiker data suffers from a calibration
uncertainty, so a small shift in zero-point could explain the deficit
in the three visible bands but this effect does not alter the slopes
of the counts.  In the data from Smail et al. (1995), the plotted
points correspond to the average of two single fields. Their V counts
are in very good agreement with ours but their R counts show a
significant number excess by a factor of about 1.2 compared with all
the others authors below R$<$ 24. \\ 
Finally we compare our deep counts with the recent bright galaxy
counts in B and R bands performed by Bertin \& Dennefeld (1996). These
counts are in good agreement with ours and we use them to normalize
the non-evolving model kindly provided by M. Fioc. Bertin et Dennefeld
(1996) suggest : $\Phi^{\star} = 0.0035 h_{50}^3 Mpc^{-3} mag^{-1}$ in
the B band (estimated at B$\sim 19^m$), $\Phi^{\star} = 0.0033
h_{50}^3 Mpc^{-3} mag^{-1}$ in R band and we adopt an intermediate
value of $\Phi^{\star} = 0.0034 h_{50}^3 Mpc^{-3} mag^{-1}$ in the V
band ($h_{50}$ is defined by $H_0 = 50 h_{50} km s^{-1}
Mpc^{-1}$). The other parameters of the luminosity function come from
Guiderdoni and Rocca-Volmerange (1990) : $\alpha \sim -1.0$ and
$M_B^{\star} \sim -20.6$.

%%%%%%%%%%%
%%%%%%%%%%%
%--------------------------------------------------------

%--------------------------------------------------------
%--------------------------------------------------------

%%%%%%%%%%%
%%%%%%%%%%%
\subsubsection{Galaxy colours}
\begin{figure*}     
\centerline{\psfig{figure=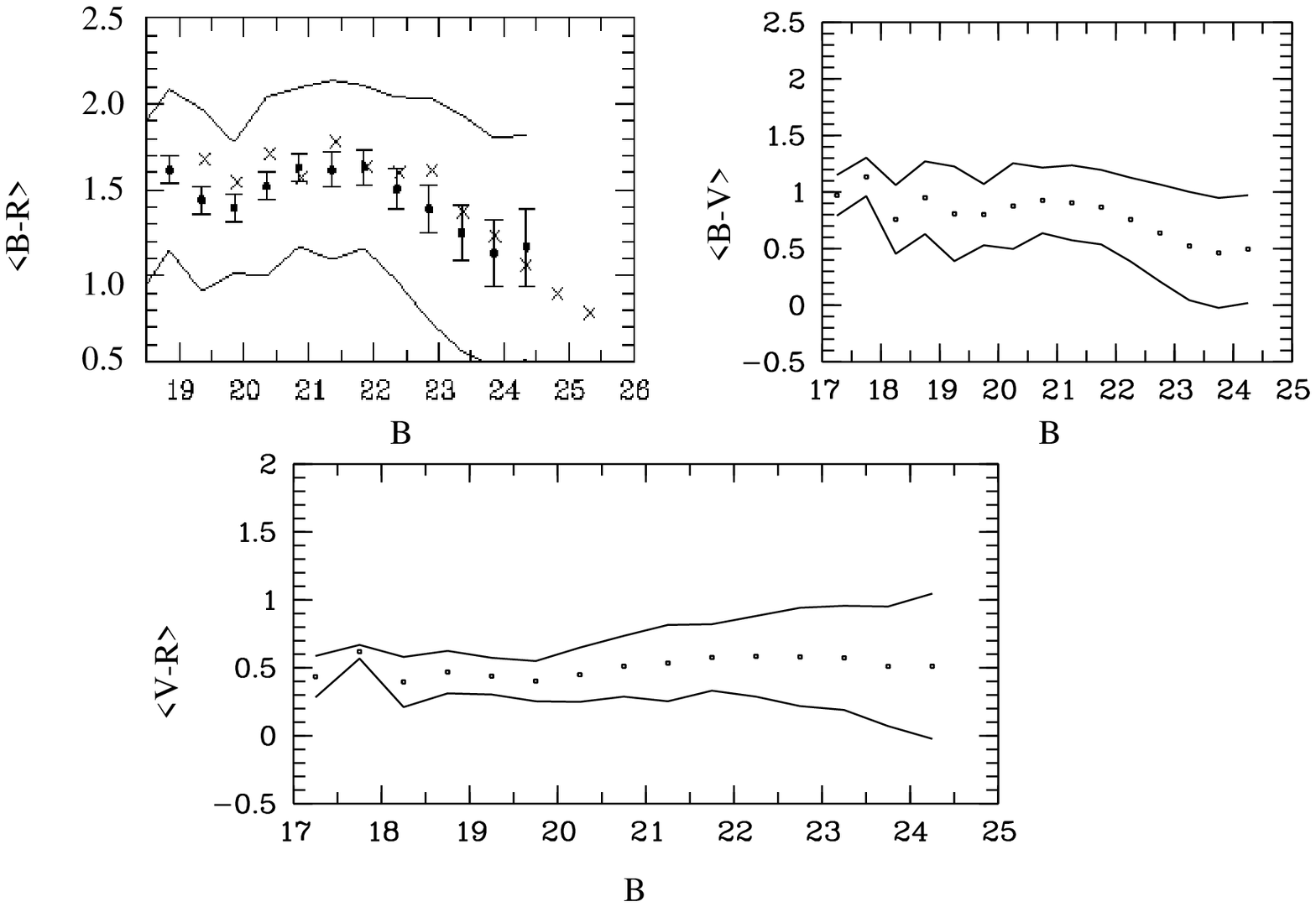,angle=-90,height=10cm,width=14.cm}}
       \caption[]{Galaxy mean colours as a function of 0.5 bin of B magnitude. 
 The solid lines represent the 1$\sigma$ envelope of the measured colours.
 The upper left graph show the mean B$-$R from our data (filled circles) and 
 from the data of Metcalfe et al.(1991,1995) (crosses). The error bars give
 the quadratic errors in the magnitudes obtained from the simulations. }
            \label{coulgalb}
    \end{figure*}

\begin{figure*}     
\centerline{\psfig{figure=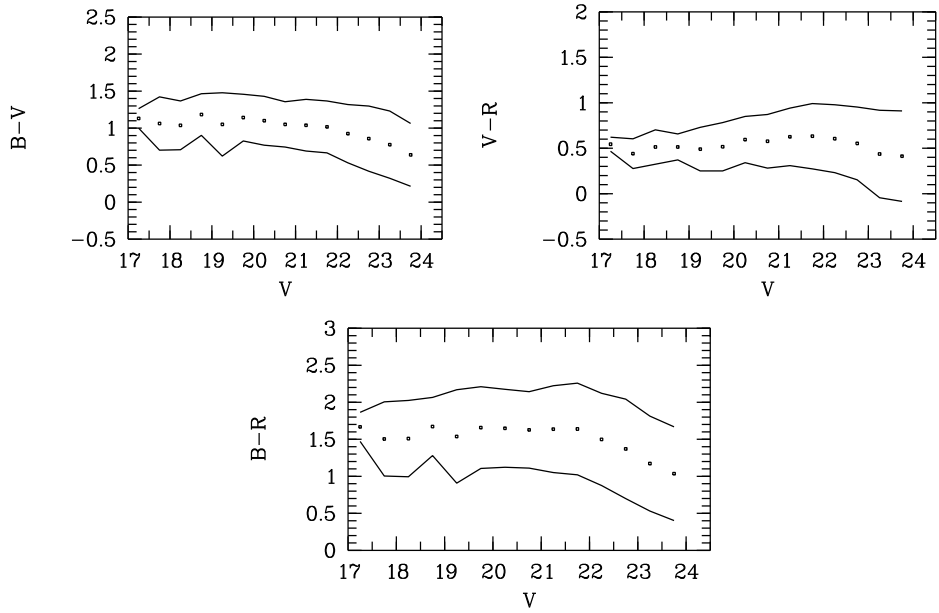,angle=-90,height=10cm,width=14.cm}}
       \caption[]{same as Fig.~\ref{coulgalb} for  V magnitude.}
            \label{coulgalv}
    \end{figure*}

\begin{figure*}     
\centerline{\psfig{figure=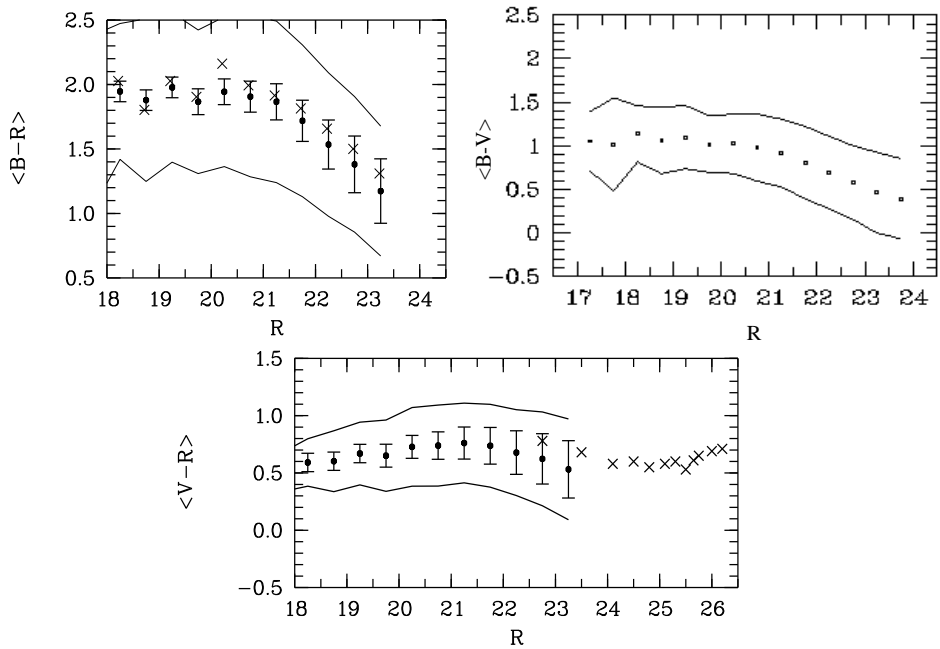,angle=-90,height=10cm,width=14.cm}}
       \caption[]{ same as Fig.~\ref{coulgalb} for R magnitude. 
 The upper left graph show the mean B$-$R from our data (filled circles) and 
 from the data of Metcalfe et al.(1991,1995) (crosses). The lower graph compare
  the mean V$-$R from our data (filled circles) with those from Smail et al.
 (1995) (crosses).}
            \label{coulgalr}
    \end{figure*}

In this section, we present the first results of the colour
distributions for our galaxy catalogue in the B, V and R bands.  As
shown in Fig.~\ref{complB}, there is a large fraction of objects
identified in all 3 filters for B$<$ 24 ($\sim$ 95 \% of objects).  In
the following, we restrict the sample to these 7500 common objects.\\
In Fig.~\ref{coulgalb},~\ref{coulgalv},~\ref{coulgalr} we plot the
mean observed colours B$-$R, B$-$V, V$-$R as a function of magnitude
for the 7500-object sample.  The solid lines draw the 1$\sigma$
envelope of the measured colours, $\sigma$ being the r.m.s. dispersion
of the colour histogram within the corresponding magnitude bin. This
envelope is larger in B$-$R than in B$-$V and V$-$R because the
expected colour in B$-$R varies in a large range for the different
galaxy types.  The B$-$R and B$-$V colours clearly show a tendency to
become bluer at fainter magnitude as first observed by Kron (1980) and
subsequently confirmed by several groups (Tyson 1988, Metcalfe et
al. 1991, 1995). \\ 
At V $\le$ 21.5, the typical mean colour is B$-$R
$\sim$ 1.55 and a blueing shift to B$-$R $\sim$ 1.0 is seen between 22
$<$ V $<$ 24.  The same tendency is visible in the B$-$V colour
distribution.  It decreases from B$-$V $\sim$ 1.05 at V $<$ 22 to
B$-$V $\sim$ 0.6 at V = 24.  Note that in Fig.~\ref{coulgalb} , at B
$>$ 24 mag, the completeness level drops to $\sim$ 60\% (see
Fig.~\ref{complB}) and the mean colours (B$-$R) and (B$-$V) become
redder due to the incompleteness in the R and V bands where only the
brighter objects are identified and contribute to shift the colour
toward redder colours. \\ 
In contrast, the (V$-$R) colour in
Fig.~\ref{coulgalb} - ~\ref{coulgalr} shows no evidence of colour
evolution with magnitude up to V $\le$ 24.  The same stability is
obtained by Driver et al. (1994).\\ 
In addition, we compare our
observed mean colours B$-$R with those from Metcalfe et
al. (1991,1995) and our mean V$-$R colours with those of Smail et
al. (1995).  Metcalfe et al's B$-$R colours are systematically 0.1 mag
redder than ours.  Half of this shift is expected due to reddening in
the fields of Metcalfe et al (Metcalfe 1995).  Comparison with the
V$-$R colours of Smail also show a small offset, but the restricted
overlap in the magnitude ranges covered does not allow us to draw any
firm conclusions about the agreement between the datasets.
%--------------------------------------------------------
%--------------------------------------------------------
%--------------------------------------------------------
%--------------------------------------------------------
%--------------------------------------------------------
%--------------------------------------------------------

\section{CONCLUSION}

We have obtained CCD photometry for 9500, 12150, 13000 galaxies in the
B, V, R (Johnson-Cousins) filters to limiting magnitudes of 24.5,
24.0, 23.5 respectively over an area of 0.4 deg$^2$ near the southern
galactic pole.  Automatic procedures were designed in order to reduce
the data in an homogeneous fashion. The main steps used in obtaining
the photometric catalogue are summarized as follows :\\
i) We have applied the standard pre-reduction techniques of bias
subtraction, flat-fielding by median filtering and cosmic-ray
removal.\\
ii) Because different CCDs were used over the course of the programme,
a large set of colour equations and zero-points were measured in order
to provide reliable calibrations for the survey. We could then
transform all instrumental magnitudes into the Johnson-Cousins
standard system.\\
iii) The survey data is a mosaic of $\simeq$ 50 CCD frames in each
band which have $\sim$ 1 arcmin overlaps.  We could therefore perform
an internal and global readjustment of the zero-point of each CCD over
the whole survey.  In this way, we reduce the zero-point
field-to-field scatter from $\sim$ 0.11 mag to $\sim$ 0.04 mag, in
agreement with the photometric uncertainty estimated using simulated
images (0.05$^m$ for R $\le$ 22 and 0.2$^m$ at fainter magnitudes).
We therefore reduce the internal dispersion in our photometry to the
intrinsic dispersion resulting from the limitations of the photometric
package. \\
iv) Accurate astrometry is performed for each CCD frame. The
positional accuracy is 0.1$\arcsec$ for objects with R$\le$ 22 and
increases to 0.3$\arcsec$ at fainter magnitudes, as estimated from
objects located in the overlapping edges of the CCDs.\\
v) Measurement of the object colours is performed by matching the
astronomical coordinates in each band with a tolerance of 1.2$\arcsec$
in position offset. The colour completeness with respect to the B band
(fraction of B objects with R and V counterparts) is greater than 95\%
up to $B \leq 24$ and drops to 60\% at 24.5. The colour uncertainties
are estimated as the corresponding quadratic sum of the errors in the
photometric magnitudes ($\sigma_{(B-V),(V-R),(B-R)} \le$0.07 for
R$\le$ 22 and $\sigma_{(B-V),(V-R),(B-R)} \le$0.28 for R$\ge$ 22). \\
vi) The star-galaxy separation is performed using a neural network and
is expected to be close to a 95\% success rate for R $\le$ 22. We use
the R band for the star/galaxy separation because it is our deeper
band, it was observed with the best seeing conditions, and it is used
for the spectroscopic selection. At R$\ge$ 22, no separation is done
due to the small fraction of faint stars expected ($\sim$ 4\%) at the
high galactic latitude of the survey ($b^{II} \sim -83^{\circ}$).

 The first results from the large photometric catalogue show that :\\
i) The galaxy counts in apparent magnitude and their slopes in
logarithmic scale are in good agreement with previous CCD and
photographic surveys and show an excess in all 3 bands with respect to
the non-evolving models. \\
ii) The (B$-$V) and (B$-$R) median galaxy colours show a blueing shift
of $\sim$0.5 mag from B $\sim$ 22 to B $\sim$24.5 . In contrast, the
(V$-$R) median colour is nearly constant up to R $\sim$ 23.5.\\
iii) The galaxy counts are well fitted by a no-evolving model for
R$\le$20.  using a $\Phi^{\star}$ derived from the new bright galaxy
counts of Bertin \& Dennefeld (1996). \\
iv) Two peaks in the stellar distributions are present in good
agreement with the Robin \& Cr\'ez\'e (1986) model of the galactic
disk and halo.\\ 
The next step in the study of this photometric sample
is the analysis of the angular correlation function $\omega(\theta)$
at faint magnitude (Arnouts \& de Lapparent 1996).  This work will
provide information about the galaxy clustering up to faint magnitudes
over a wide range of scales.  The high completeness rate in colour
should also allow to further characterize the properties of the faint
blue galaxies.\\ 
Together with the photometric data, the redshift survey of $\sim$ 700
galaxies with R $\le$ 20.5 will provide a multi-colour optical
luminosity function in the redshift range 0.1 $\le z \le$ 0.5.
Altogether, these data will allow us to address the issue of the
evolution of galaxies with B $\le$ 22. The photometric and redshift
catalogues will also be complemented by a spectral classification of
all galaxies with R$\le$20.5 (Galaz \& de Lapparent 1996) and will
thus provide an unique database for studying the variations in the
galaxy properties as a fonction of environment and redshift.
\begin{acknowledgements}

We are grateful to the European Southern Observatory for the large
amount of observing time and the corresponding logistic support which
allowed to perform this observing programme in good conditions and to
bring it to completion. We are grateful to Dr P. Leisy for kindly
providing us his cosmic removal algorithm, and to M. Fioc and Dr
B. Rocca-Volmerange for kindly providing us the non-evolving model to
fit the differential galaxy number counts.  We thank Dr C. Willmer for
fruitful discussions. We also thank the referee Dr. N. Metcalfe for
his scientific comments and linguistic corrections.
\end{acknowledgements}  


\begin{thebibliography}{}
%   \bibitem{} Babul A. and Rees M., 1992, MNRAS 255, 346 
   \bibitem{} Arnouts S., de Lapparent V., 1996,  in preparation
   \bibitem{} Bellanger C., de Lapparent V. and Arnouts S. et al., 1995a, A\&A Sup. Ser. 110, 159    
   \bibitem{} Bellanger C. et de Lapparent V. 1995b, ApJ L., 455, L103  
   \bibitem{} Bertin E., 1996, PhD thesis, IAP
   \bibitem{} Bertin E. and Arnouts S. (BA96), 1996, A\&A Sup Ser 117, 393 
   \bibitem{} Bertin, E. and Dennefeld, M., 1996, accepted for publication in A\&A.
   \bibitem{} Broadhurst T.J., Ellis R.S. and Shanks T., 1988, MNRAS 235, 827 
   \bibitem{} Broadhurst T.J., Ellis R.S., Koo D.C. and Szalay A.S., 1990, Nature
343, 396
   \bibitem{} Broadhurst  T.J., Ellis R.S. and Glazebrook K., 1992, Nature 355, 55
   \bibitem{} Bruzual A.G., 1983, ApJ 273, 105 
   \bibitem{} Burki G., Rufener F., Burnet M. et al., 1995, A\&A Supp Ser 112, 383 
   \bibitem{} Buzzoni B., Delabre B., Dekker H. et al., 1984, The Messenger 38, 9
   \bibitem{} Colless M. M., Ellis R. S., Taylor K., Hook R. N., 1990, MNRAS 244,
408 
   \bibitem{} Colless M. M., Ellis R. S., Broadhurst T.J., Taylor K., Peterson
B.A. 1993, MNRAS 261, 19
   \bibitem{} Cowie L.L., Songaila A., Hue E.M., 1991, Nature 354, 460
   \bibitem{} Dalcanton J.J., 1993, ApJ 415, L87
   \bibitem{} da Costa L.N., Geller M.J., Pelligrini P.S. et al 1994, ApJ 424, L1
   \bibitem{} Dekker H., d'Odorico S., Kotzlowski H. et al. 1991, The Messenger 63,
73
   \bibitem{} de Lapparent V., Geller M.J., Huchra J.P., 1986, ApJ 302, L1
   \bibitem{} de Lapparent V., Geller M.J., Huchra J.P., 1991, ApJ 369, 273
   \bibitem{} d'Odorico S. 1990, The Messenger 61, 51
   \bibitem{} Driver S.P., Phillipps S., Davies J.I., Morgan I., Disney M.J., 1994, MNRAS 266, 155
%   \bibitem{} Driver S.P., Phillipps S., Davies J.I., Morgan I., Disney M.J., 1994, MNRAS 268, 393 
%   \bibitem{} Efstathiou G., Ellis R.S., Peterson B.A., 1988, MNRAS 232, 4311988 
   \bibitem{} Efstathiou G., Bernstein G., Katz N., Tyson J.A., Guhathakurta P., 1991, ApJ 380, L47 
%   \bibitem{} Fukugita M., Takahara F., Yamashita K., Yoshii Y., 1990, ApJ 361, L1  
   \bibitem{} Geller M.J. and Huchra J.P. 1989, Science 246, 897
   \bibitem{} Guiderdoni B. and Rocca-Volmerange B., 1990, A\&A 227, 362
   \bibitem{} Glazebrook K., Ellis R., Colless M. et al. 1995a, MNRAS 273, 157
   \bibitem{} Glazebrook K., Ellis R., Santiago B. and Griffiths R., 
1995b, MNRAS, preprint
   \bibitem{} Graham J., 1981, PASP 93, 29
   \bibitem{} Infante L., 1986, PASP 98, 360
   \bibitem{} Koo D.C. and Kron R.G., 1992, Ann. Rev. Astr. Astrophys., 30, 613
   \bibitem{} Koo D.C., Gronwall C., Bruzual G.A., 1993, ApJ 415, L21
   \bibitem{} Kron R.G., 1980, ApJS 43, 305
   \bibitem{} Landolt A.U., 1992, AJ 104, 340  
   \bibitem{} Leisy, P., 1994, Private communication
   \bibitem{} Lilly S.J., Cowie L.L. and Gardner J.P., 1991, ApJ 369, 79
   \bibitem{} Majewski S. R., 1992, ApJSS 78, 87
   \bibitem{} Maddox S.J., Efstathiou G., Sutherland W.J., 1990a, MNRAS 246, 433
   \bibitem{} Maddox S.J., Sutherland W.J.,  Efstathiou G., Loveday J., Peterson
B.A., 1990b, MNRAS 247, 1p 
   \bibitem{} Metcalfe N., Fong R., Shanks T., Jones L. R., 1991, MNRAS 249, 498
   \bibitem{} Metcalfe N., Fong R., Shanks T., Jones L. R., 1995, MNRAS 
   \bibitem{} Metcalfe N., 1995, Private communication
   \bibitem{} Moffat A.F.J., 1969, A\&A, 3, 455 
   \bibitem{} Ostriker J.P., 1990, in The Evolution of the Universe of Galaxies, ed.
R. Kron, 25
   \bibitem{} Revenu B., de Lapparent V., 1992, private communication
   \bibitem{} Robin and Cr\'ez\'e, 1986, A\&A 157, 71
   \bibitem{} Roeser, S. and Bastian, U., 1991, PPM Star Catalogue, Spektrum
Akademisher Verlag, Heidelberg, Berlin, New York 
   \bibitem{} Ramella M., Geller M.J. and Huchra J.P., 1992, ApJ 384, 396
   \bibitem{} Roche N., Shanks T., Metcalfe N., and Fong R., 1993, MNRAS 263, 360 
   \bibitem{} Rocca-Volmerange B. and Guiderdoni B., 1990, MNRAS 247, 166 
   \bibitem{} Schwarz H. E. and Melnick J., 1993, The ESO Users Manual 
   \bibitem{} Seldner M., Siebers B., Groth E. J., Peebles P.J.E.,1977,  AJ 82, 249  
   \bibitem{} Shanks T., Phillipps S., Fong R., 1980, MNRAS 191, 47p.
   \bibitem{} Shanks T., Stevenson P.R.F., Fong R., McGillivray H.T., 1984, MNRAS
206, 767
   \bibitem{} Smail I., Hogg D.W., Yan L., Cohen J.G., 1995, preprint  
   \bibitem{} Tresse L., Hammer F., Le f\`evre O., Proust D., 1993, A\&A, 277, 53
   \bibitem{} Tyson J.A., 1988, Astr. J. 96, 1.
   \bibitem{} West R., Kruszewski A., 1981, Irish Astron. J. 15, 25. 
   \bibitem{} Yoshii Y. and Takahara F., 1988, ApJ 326, 1

\end{thebibliography}
\end{document}